\documentclass[aps,pre,showpacs]{revtex4-1}

\usepackage{graphicx}
\usepackage{dcolumn}
\usepackage{bm}

\begin{document}


\title[Multifractal surrogate data with preserved regularity]{A multifractal surrogate data generation algorithm that preserves pointwise H\"{o}lder regularity structure, with initial applications to turbulence}

\author{C. J. Keylock}
\affiliation{Sheffield Fluid Mechanics Group and Department of Civil and Structural Engineering, University of Sheffield, Mappin Street, Sheffield, U.K., S1 3JD }%
 \email{c.keylock@sheffield.ac.uk}

\date{\today}

\begin{abstract}
An algorithm is described that can generate random variants of a time series or image while preserving the probability distribution of original values and the pointwise H\"{o}lder regularity. Thus, it preserves the multifractal properties of the data. Our algorithm is similar in principle to well-known algorithms based on the preservation of the Fourier amplitude spectrum and original values of a time series. However, it is underpinned by a dual-tree complex wavelet transform rather than a Fourier transform. Our method, which we term the Iterated Amplitude Adjusted Wavelet Transform (IAAWT) method can be used to generate bootstrapped versions of multifractal data and, because it preserves the pointwise H\"{o}lder regularity but not the local H\"{o}lder regularity, it can be used to test hypotheses concerning the presence of oscillating singularities in a time series, an important feature of turbulence and econophysics data. Because the locations of the data values are randomized with respect to the multifractal structure, hypotheses about their mutual coupling can be tested, which is important for the velocity-intermittency structure of turbulence and self-regulating processes. 
\end{abstract}

\pacs{05.45.Tp,05.45.Df, 02.50.Fz, 47.27.-i, 07.05.Pj}
\keywords{multifractality, surrogate data, turbulence, H\"{o}lder regularity}
\maketitle

\section{\label{sec:level1}Introduction}
Given a measurement of a complex, and potentially nonlinear phenomenon in the form of an image or a time series, it is often important and insightful to test the statistical significance of metrics that attempt to characterize and summarize the underlying data structure. Because a complex process does not necessarily have a Gaussian or even a known prior distribution function, conventional statistical analyses are not always readily applicable. Hence, the surrogate data approach was developed within nonlinear physics in the 1990s as a means to treat this class of problems \citep{T92,PT94,ss96,small01,k06}. This has proven to be a popular methodology, with applications spanning many disciplines, including medical physics \cite{theiler96,govindan98,ivanov99}, fluid mechanics \citep{basu04,poggi04} and the geosciences \cite{volobuev08,k14b}.

Perhaps the best known algorithm is the Iterated Amplitude Adjusted Fourier Transform (IAAFT) method \citep{ss96}, which gives surrogates constrained to the original values in the data, while also preserving the Fourier amplitude spectrum accurately. Consequently, differences between the data and surrogates will reside in a specific structure to the Fourier phases of the original data that cannot be replicated by chance in the synthetic surrogate data. Consequently, hypotheses concerning the possible nonlinear properties of an observed time series may be answered by comparing the values of some metric that characterises the nonlinearity (e.g. the correlation dimension, maximal Lyapunov exponent, derivative skewness, nonlinear prediction error) \citep{ss97} between the data and the surrogates. Thus, having selected a statistical significance level $\alpha$ and assuming a two-tailed statistical test, so that $\frac{2}{\alpha}-1$ surrogates are generated, if the value of the metric for the data falls outside the range for the surrogates then a significant difference exists.

A development of this general approach is termed Gradual Wavelet Reconstruction (GWR) and envisages the original data to be at one end of a continuum ($\rho =1$) and phase-randomised, IAAFT data to be at the other ($\rho = 0$). Surrogate data along this continuum are generated by fixing in place some of the phase information in the original time series \cite{k10}. This is achieved by adopting a wavelet transform and, indeed, wavelets and windowed Fourier transform methods have been used in the past as a means to generate IAAFT-like surrogates \citep{B03} and for constraining the randomization in other ways \citep{k07,flandrin09}. With surrogates generated at different values for $\rho$ it is then possible to generate bootstrapped confidence limits for data that preserve a sufficient degree of the nonlinearity present in the time series \citep{k11,k12c}, or to test relative degrees of complexity on a given metric by determining the value for $\rho$ above which there is no longer a significant difference between data and surrogates \citep{k14}.   

Given that one way in which a time series may depart from its IAAFT surrogates is through the presence of multifractality, a different approach to GWR for this case is to construct surrogates that preserve the multifractal characteristics of a time series \citep{PALUS}. Such surrogates can be used to generate bootstrapped variants of multifractal phenomena such as human cardiac dynamics \citep{ivanov99}, earth surface topography \citep{GLS03}, precipitation \citep{ven06}, stock market price fluctuations \citep{MATIA}, or Quantum Hall systems \citep{KM95}. They may also be used as a means for testing if the observed degree of multifractality departs significantly from that of a constant pointwise H\"{o}lder exponent. However, the existing multifractal algorithm of \citet{PALUS} does not impose the additional constraint that the values of the original dataset is preserved, making the algorithm more akin to the original surrogate data methods of Theiler and co-workers \citep{T92,PT94} rather than the IAAFT algorithm, where differences have to reside in the phase information. 

This paper gives a new algorithm that not only incorporates this feature, but also approximates the ordering in time (or space for the image processing variant) of the H\"{o}lder exponents for the original data. We demonstrate the utility of the method by applying it to the detection of oscillating singularities in time series \citep{jaffard96,arneodo98}, which is relevant to furthering our understanding of turbulence \citep{hunt91,nicolleau99} and to analysing turbulent systems where there is a dependence between the intermittency and the macroscale velocity, as originally discussed by Kolmogorov \citep{K62}, and subsequently shown to be of relevance in various contexts \citep{prask93,stresing10,k12}. Before the new algorithm is explained and applied, we briefly review essential concepts of multifractality and H\"{o}lder exponents, leading to methods for their estimation. We then review surrogate algorithms for IAAFT surrogate generation as well as the method introduced by \citet{PALUS}. After presentation of our method, we test its relative effectiveness and then provide some example applications of relevance to turbulence physics.

\section{H\"{o}lder regularity}
\subsection{Definition of pointwise and local H\"{o}lder regularity}
In this paper we primarily focus on the pointwise H\"{o}lder regularity, $\alpha_{p}(t)$, of a function, the variation in which is approximated by methods for characterizing multifractality such as structure functions \citep{fp85,ess}, universal multifractal exponents \citep{schertzer87} or wavelet transform modulus maxima methods \citep{MUZY91,BMA93}. However, in addition, one may define the local H\"{o}lder regularity, $\alpha_{L}(t)$ to characterize oscillating singularities \citep{SL02,KL02}, which are not distinguished from sharp singularities when using $\alpha_{p}(t)$, and can go further and define a function of such exponents calculating all forms of regularity at a point \citep{KL02}. We make use of $\alpha_{L}(t)$ when looking at the statistical detection of an oscillating singularity, as described below. 

Given a signal, $u_{t}$, a position, $t_{0}$, and assuming that $\alpha_{p} \in \{0,\ldots,1 \}$ then we seek values, $\beta_{p}$, for which
\begin{equation}
| u(t) - u(t_{0})| \le c_{p} |t - t_{0}|^{\beta_{p}}
\label{eq.pH}
\end{equation} 
The value for $\alpha_{p}(t_{0})$ is then the supremum of this set of legitimate $\beta_{p}$ values. Following \citet{SL02}, to define $\alpha_{L}$, we commence by considering a ball, $B(t_{0},R)$, with a centre at $t_{0}$ and with radius, $R$. If the constant, $c_{L}$, exists such that the following inequality holds:
\begin{equation}
| u(t_{i}) - u(t_{j})| \le c_{L} |t_{i} - t_{j}|^{\beta_{L}}
\label{eq.pL}
\end{equation} 
where $t_{i},t_{j}$ ($i \ne j$) fall within $B(t_{0},R)$, then the choice for $\beta_{L}$ is legitimate and $\alpha_{L}(t_{0},B(t_{0},R)$ is given by the supremum of these $\beta_{L}$. Finally, the local H\"{o}lder exponent, $\alpha_{L}(t_{0})$ is given by
\begin{equation}
\alpha_{L}(t_{0}) = \lim_{R \to 0} \alpha_{L}(t_{0},B(t_{0},R))
\end{equation}
and $\alpha_{L} \le \alpha_{p}$. 

This formulation permits us to contrast two functions - a cusp and a chirp - or in the terminology of \citet{hunt91}, a point vortex and a spiral vortex:
\begin{eqnarray}
\textrm{Cusp:  }u(t) &=& |t-t_{0}|^{\alpha_{+}} \\
\textrm{Chirp:  }u(t) &=& |t-t_{0}|^{\alpha_{+}} \sin \frac{1}{|t-t_{0}|^{\alpha_{\omega}}}
\label{eq.chirp} 
\end{eqnarray}
In the former case, at $t = 0$, $\alpha_{p} = \alpha_{L} \equiv \alpha_{+}$. In the latter, $\alpha_{p} = \alpha_{+}$ and $\alpha_{\omega} = (\alpha_{p} / \alpha_{L}) - 1$. Hence, $\alpha_{p}$ measures the density of singularities, but is insensitive to their type. On the other hand, $\alpha_{L}$ provides information on the topology of these singularities.

\subsection{Combined estimation of $\alpha_{p}$ and $\alpha_{L}$ in a function space}
There are a number of techniques that exist for estimating $\alpha_{p}(t)$ \citep{KL02} and $\alpha_{L}(t)$ \citep{arneodo98,jaffard96}, but if both are to be determined simultaneously it is important that the methods used are consistent. This can be assured using a time domain, microlocal frontier method \citep{KL02,SLV03} that resolves both $\alpha_{p}(t_{0})$ and $\alpha_{L}(t_{0})$ simultaneously in the time domain. A function is said to belong to $K_{t_{0}}^{(s, s^{'})}$, if for $0 < \delta < 1/4$, $c_{k} > 0$, and for all $(t_{i}, t_{j})$ where $|t_{i} - t_{0}| < \delta$ and $|t_{j} - t_{0}| < \delta$, the following is true:
\begin{eqnarray}
| u(t_{i}) - u(t_{j})| &\le& c_{K} |t_{i} - t_{j}|^{s + s^{'}}(|t_{i} - t_{j}| + |t_{i} - t_{0}|)^{-s^{'}/2} \nonumber \\
&\times& (|t_{i} - t_{j}| + |t_{j} - t_{0}|)^{-s^{'}/2}
\label{eq.front}
\end{eqnarray}
A comparison to (\ref{eq.pH}) and (\ref{eq.pL}) shows that $K_{t_{0}}^{(s, s^{'})}$ includes the definitions for both $\alpha_{p}$ and $\alpha_{L}$. Setting $s^{'} = 0$ recovers (\ref{eq.pL}) and in the limit $\delta \to 0$, one obtains $s \le \alpha_{L}(t_{0})$, meaning that the supremum is $\alpha_{L}(t_{0})$. Because the method is concerned with first order differences (increment statistics), both $s$ and $-s^{'}$ have a maximum of 1. (Note that higher order singularities can be studied by replacing the left hand side of (\ref{eq.front}) with a higher order difference \citep{SLV03}). Hence, it follows that $s + s^{'} = 1$ is an upper bound on the meaningful behavior of $K^{(s, s^{'})}$. Forming the ratio between the left hand side of (\ref{eq.front}) as the numerator and the right hand side as the denominator, it is clear that this ratio tends to $\infty$ if $s + s^{'} > \alpha_{p}(t_{j})$ for fixed $t_{j}$ and increasing $t_{i}$. Given that $\alpha_{L}(t_{0}) \le \alpha_{p}(t_{0})$ it follows that the value for $s$ that marks the intersection between values on the convex frontier for $K_{t_{0}}^{(s, s^{'})}$ and $s + s^{'} = 0$ gives $\alpha_{p}(t_{0})$. Similarly, the value for $s$ where $K_{t_{0}}^{(s, s^{'})}$ intersects $s^{'} = 0$ gives $\alpha_{L}(t_{0})$ \citep{KL02}. This is the method adopted here for the simultaneous estimation of the pointwise and local exponents \citep{fraclab}. 

\section{Surrogate data algorithms}
Both the IAAFT method and the multifractal method of \citet{PALUS} are relevant to our approach. The IAAFT algorithm for a discrete time series $x_{t}$, $t = 1, \ldots, N$ may be stated as \citep{ss96}:
\begin{enumerate}
\item Store the squared amplitudes of the discrete Fourier transform of $x_{t}$ (i.e. $X_{f}^{2} = |\,\sum^{N}_{1}x_{t}\,e^{i2\pi\,f(t/N)}|^{2}$);
\item perform a random shuffle of $x_{t}$ to give $x_{t}^{(j=0)}$;
\item Then iterate a power spectrum step and a rank-order matching step on $x_{t}^{(j)}$ as follows:
\begin{enumerate}
\item Take the Fourier transform of $x_{t}^{(j)}$ and replace the squared amplitudes with $X_{f}^{2}$, while retaining the phases. Given the initial random sort, this means that the spectrum should be preserved but with random phases. Invert the Fourier transform with the original amplitudes restored;
\item Replace the values in the new series $x_{t}^{(j)}$ by those in $x_{t}$ using a rank-order matching process. This preserves the set of original values in the dataset but deteriorates the quality of spectral matching, which explains why the Fourier amplitudes are only replicated approximately;
\end{enumerate}
\item Repeat until a convergence criterion is fulfilled or any changes are too small to result in any re-ordering of the values from the previous iteration.
\end{enumerate}
In this way, the histogram of the original data is preserved precisely, and the Fourier spectrum is approximated to a given error tolerance.
 
The multifractal approach \citet{PALUS} is conceptually similar to a discrete wavelet transform (DWT) based algorithm for realising multifractal data \citep{B93,ABM98}. Synthetic wavelet coefficients are generated on a dyadic tree that, when inverted yields a dataset with prescribed multifractal characteristics. If the chosen distribution function, $P(\eta)$, is symmetric about zero then, given a value for the wavelet coefficient at the largest scale, $w_{0,0}$, the coefficients at scale, $j$, and position, $k$, are found by:
\begin{equation}
\label{eq.1}
 w_{j,k}=\eta_{j,k} \, w_{j-1,\frac{1}{2}k}
\end{equation}
If $P(\eta)$, is asymmetric about zero, then this is modified to
\begin{equation}
\label{eq.2}
 w_{j,k}=\epsilon_{j,k} \, \eta_{j,k} \, w_{j-1,\frac{1}{2}k}
\end{equation}  
where, with equal probability, $\epsilon_{j,k} = \pm 1$. The \citet{PALUS} algorithm commences with a DWT of a dataset and, with scales $j \in \{0,1\}$ left fixed, then builds surrogate values for $\eta_{j,k}$ for $j > 1$, designated here by $\tilde{\eta}_{j,k}$ according to:
\begin{equation}
\tilde{\eta}_{j,k} = \frac{w_{j,k}}{w_{j-1,\frac{1}{2}k}} \,\,\,\,\,\,\,\,\,\,\,\,j > 1
\end{equation}
At each scale, $j$, the $2^{j}$ multipliers $\tilde{\eta}_{j,k}$ are then permuted to give values  $\tilde{\mu}_{j,k}$. Setting $\tilde{w}_{j,k} = w_{j,k}$ for $j \in \{0,1\}$, the new wavelet coefficients are constructed for $j>1$ by:
\begin{equation}
\tilde{w}_{j,k} = \tilde{\mu}_{j,k} \, \tilde{w}_{j-1,\frac{1}{2}k}
\end{equation}
Amplitude adjustment is then used to recover the original distribution of wavelet coefficients at each scale. That is, at a scale, $c$, the $\tilde{w}_{j=c,k}$ are replaced by the equivalent ranked values from $w_{j=c,k}$. In this way, the permutations of the DWT are randomized in a manner that preserves the original hierachical structure. From this randomized dyadic structure of the $w_{j,k}$,  the inverse DWT can be employed to generate a new data sequence. Note that the amplitude adjustment step of this algorithm is applied to the $w_{j,k}$. Consequently, the histogram of the new data sequence will not exactly replicate the values for the original time series.

\begin{figure}[t]
\vspace*{2mm}
\begin{center}
\includegraphics[width=11.6cm]{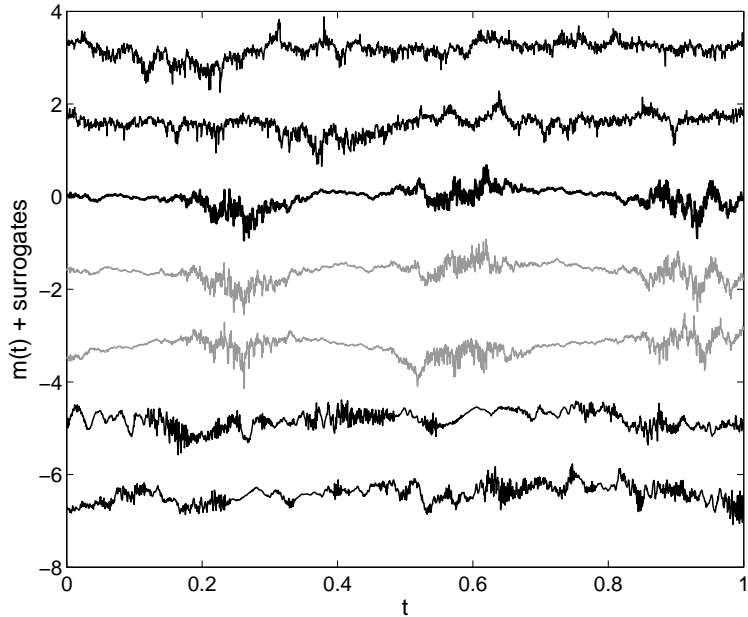}
\end{center}
\caption{A multifractional Brownian motion, $m(t)$, is shown as a heavy black line at the origin and is generated from $\alpha_{p}(t) = 0.33 + 0.2 sin (6 \pi t)$. Two example IAAFT surrogates are shown in black and displaced vertically above $m(t)$, while two example Palu\u{s} surrogates are in black and displaced below $m(t)$. Surrogates generated by the new algorithm are shown in gray and immediately below $m(t)$.}
\label{fig.1}
\end{figure}

Figure ~\ref{fig.1} shows an example multifractional Brownian motion \citep{benassi97,peltier95}, $m(t)$, generated from an underlying sinusoidal function for the pointwise H\"{o}lder regularity, $\alpha_{p}(t)$, using the algorithm of \citet{woodchan}: $\alpha_{p}(t) = 0.33 + 0.2 \sin(6\pi t)$ as the black solid line together with two example surrogates from the IAAFT algorithm (black and displaced upwards) and using the Palu\u{s} algorithm (black and displaced downwards). The IAAFT algorithm returns the correct average roughness because of the Fourier spectrum preservation. However, it clearly fails to capture the intermittency in the test signal. The Palu\u{s} algorithm does a better job in this respect, but it is clear visually that the superior surrogates are those in gray, which are produced using the algorithm presented in this paper. In addition, while the data values in these latter surrogates occur at random locations, the H\"{o}lder properties are clearly localised - the rough and smooth patches occur at the same positions in the data and surrogates. This is a desired property of the new algorithm as shown in the example applications, below. 

\section{The IAAWT algorithm}
A pair of dyadic wavelet trees may be designed to form a Hilbert transform pair \citep{seles02} and such wavelets have been proposed for analysis of turbulence \citep{af94} and waveform encoding \citep{ozturk}. More specifically, we employ the dual-tree complex DWT \citep{kings01,seles05}. The analytic signal of some real signal, $x(t)$ is given by $x_{a}(t)=x(t)+i\,x_{H}(t)$, where $x_{H}(t)$ is the Hilbert transform of $x(t)$, which is a convolution operator with a filter given by $h(t) = 1/(\pi\,t)$:
\begin{equation}
x_{H}(t)=\int^{\infty}_{-\infty}h(\tau)\,x(t-\tau)d\tau
\end{equation}
Because the Fourier transform of $h(t$) lies completely in the imaginary plane, it follows that a Hilbert transform approach can be used to perform a complex-valued wavelet transform. Given two filters $g(t)$ and $h(t)$ and their Fourier transforms $G(\omega)$ and $H(\omega)$ then it may be shown that if $G(\omega)=H(\omega)e^{-i\,\omega/2}$ for $|\omega| < \pi$ then their associated wavelets form a Hilbert pair \citep{seles01}, which can be achieved for orthogonal wavelets by offsetting the scaling filters by one half sample. 

Hence, one approach would be to deploy two trees of linear phase filters, of even length in one tree and odd in the other \citep{kings01}. However, such filters lack orthogonality and the sub-sampling structure is not particularly symmetric. Thus, \citet{kings01} proposed the \emph{Q-shift} dual tree where, below the coarsest scale, all filters are even length, but no longer linear in phase. By designing the filters to have a delay of $\frac{1}{4}$ sample and by using the time reverse of one set of filters in the other tree, the required $\frac{1}{2}$ sample delay can be achieved. In this paper we use symmetric, biothogonal filters with support widths of 13 and 19 values for the first level of the algorithm and \emph{Q-shift} filters with a support of 14 values for all other levels on the dual tree (case C in \citet{kings01}). The \emph{Q-shift} dual tree approach retains properties that make undecimated transforms advantageous for use in surrogate generation, such as shift invariance \citep{k06}, but at a computational cost that is merely double that for a standard discrete wavelet transform, O($N$), rather than O$(N \log_{2}N)$, which is the case for an undecimated maximal overlap discrete wavelet transform \citep{lp96}.

Our algorithm may now be stated in a form that parallels the IAAFT. Given a signal of length $N = 2^{J}$ we proceed as:
\begin{enumerate}
\item Perform a dual-tree complex DWT and obtain the amplitudes and phases over all $J$ scales for the $k =1, \ldots, 2^{j-1}$ complex valued $w_{j,k}$ at each $j$;
\item Randomly sort the original data and take the dual-tree complex DWT to produce randomised wavelet phases for each scale;
\item Produce new $w_{j,k}$ by combining the original amplitudes with the randomised phases;
\item Iterate the following steps until convergence is achieved:
\begin{enumerate}
\item Perform the inverse wavelet transform to give a new time series and then apply the same amplitude adjustment step used in the IAAFT algorithm;
\item Take the dual-tree complex DWT and obtain the new phases, combine these with the original amplitudes to give the latest $w_{j,k}$.
\end{enumerate} 
\end{enumerate}
As with the IAAFT algorithm, this new method, which we term the Iterated Amplitude Adjusted Wavelet Transform (IAAWT), preserves the probability distribution of original values in the dataset, with the multifractal structure preserved up to a convergence criterion.  

\section{Testing the IAAWT algorithm}
\subsection{Tests using signals with prescribed regularity structure}
Figure ~\ref{fig.1} shows that the IAAWT algorithm replicates the multifractal characteristics of a signal when the $\alpha_{p}(t)$ are given by a continuous and differentiable function of time. A more complex multifractal signal has its $\alpha_{p}(t)$ prescribed by a continuous and non-differentiable function, such as the Weierstrass function \citep{falconer}, or by some hierarchical process. We consider the latter by generating a signal using the algorithm of \citet{B93}. This provides a stringent test of the IAAWT algorithm because the similarities between the \citet{B93} and \citet{PALUS} algorithms imply that the latter will yield better results than the IAAWT technique. 

With reference to eq.~(\ref{eq.1}) and ~(\ref{eq.2}), we use an empirical, discrete distribution function for $P(\eta)$ where $\eta \in \{0.4, 0.5, 0.6, 0.7\}$ and the corresponding probabilities are $P(\eta) \in \{0.45, 0.1, 0.1, 0.55\}$, and with $w_{0,0}$ selected randomly from this distribution, we then apply ~(\ref{eq.2}) to derive a dyadic tree of wavelet coefficients over 16 scales, which we then invert with a least asymmetric Daubechies wavelet with 3 vanishing moments to produce a signal, $B(t)$ containing $2^{16}$ values. Nineteen further such signals were generated to characterise the inherent variability in the multifractal properties of stochastic signals with the same $P(\eta)$. The multifractal properties of these data were determined using the structure function approach \citep{fp85,ess} where, for a choice of moment exponent, $n$, the structure function is $S_{n}(r) = \langle [B(t+r)-B(t)]^{n} \rangle$ and the scaling exponent, $\xi(n)$ describes the power-law relation between $r$ and $S_{n}(r)$ \citep{fp85}. 
 
\begin{figure}[t]
\vspace*{2mm}
\begin{center}
\includegraphics[width=11.6cm]{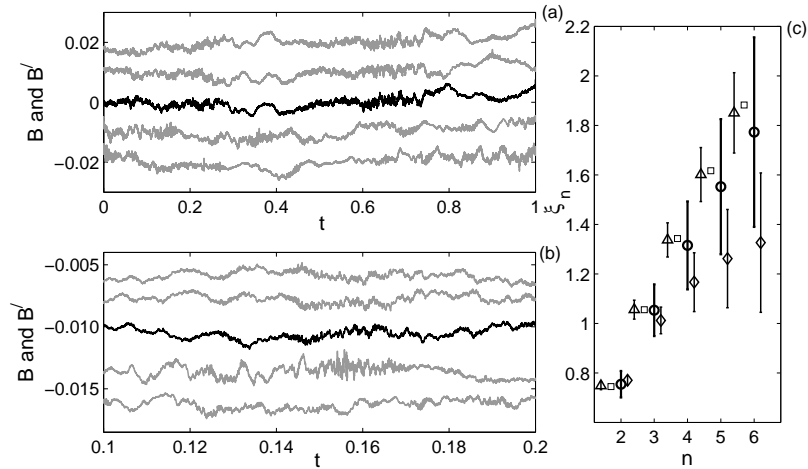}
\end{center}
\caption{A multifractal signal, $B$, generated using the \citet{B93} algorithm is shown in (a) as a black line. Panel (b) zooms in on a portion of $B$. Surrogate series generated using the IAAWT algorithm are shown in gray and displaced upwards and those from the \citet{PALUS} algorithm are in gray and displaced downwards in these two figures. Panel (c) shows structure function exponents, $\xi_{n}$ as a function of moment number, $n$, with error bars indicating $\pm$ 2 standard deviations. The dark lines with circles are for 20 different realisations of the process with constant $P(\eta)$. The squares give the structure function scaling for the data in (a) and (c), while triangles are the results for 25 IAAWT surrogates of this data series, and diamonds are the results for 25 Palu\u{s} surrogates. Values are displaced horizontally from the integer value of $n$ for clarity.}
\label{fig.2}
\end{figure}

Figure ~\ref{fig.2} shows an example signal $B$ in black in (a) and (b) together with surrogate series, $B^{/}$ in gray. Panel (b) is simply a magnification of a relatively rough region of (a) showing that this section of the signal also contains regions of relatively high and low roughness, demonstrating the hierarchical nature of the multifractality of $B$. Note that both surrogate algorithms yield multifractal data but the enhanced ability of the IAAWT algorithm to preserve the multifractal properties of the original data is quantified in (c), where the variability for all 20 different variants of $B$ are shown by the circles (median value) and heavy black lines ($\pm 2$ standard deviations). The structure function results for the data shown in Fig.~\ref{fig.2}a are given by a square and structure function estimates for 25 surrogates for this signal using the IAAWT algorithm are shown by the triangle and error bars, and using the Palu\u{s} algorithm, by a diamond with error bars. Not only is the median for the IAAWT surrogates very close to the value for the data (shown as the square), the error in estimating $\xi_{n}$ using the IAAWT is much less than that inherent to different realisations of $B$. The Palu\u{s} algorithm is both less accurate and less precise than our new method.

\begin{figure}[t]
\vspace*{2mm}
\begin{center}
\includegraphics[width=11.6cm]{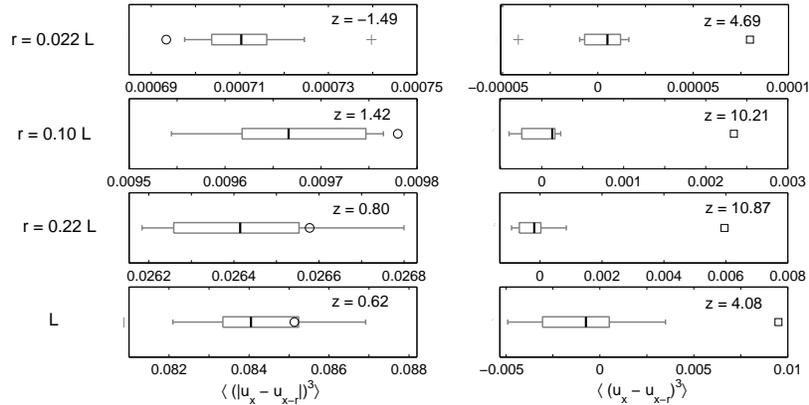}
\end{center}
\caption{An analysis of the third order moment of the velocity differences for a turbulence time series of the longitudinal velocity component, $u_{x}$. The values for $r$ range from the Taylor scale $\sim 0.02 L$ to the integral scale, $L$. The circles and squares are the values for the data, while the boxplots indicate the values for the surrogates: the median by a heavy vertical lines, the edges of the box by the lower and upper quartile, the whiskers extend up to $\times 1.5$ the interquartile range, with outliers shown by crosses.}
\label{fig.3}
\end{figure}
\subsection{Testing that characteristics other than multifractality are not preserved}
To highlight that our algorithm is only preserving the multifractal structure, we examine the ability to preserve the skewness of the increments and skewness of the absolute part of the velocity increments for a turbulence time series in ~\ref{fig.3}. The data series consists of $2^{17}$ samples of the longitudinal velocity component, $u_{x}$, measured 0.02 m ($\sim 150$ wall units above the bed of a wind tunnel in a developed boundary layer with a mean flow of 6 m s$^{-1}$, with more detail on the experimental design in the original references \cite{k12b}). Results are presented for choices of the separation $r$ that vary from the Taylor scale to the integral scale, $L$. Ten IAAWT surrogates were generated to form the boxplots in ~\ref{fig.3}. Because it is $\langle (| u_{x} - u_{x+r}|)^{3}\rangle$ (circles), rather than $\langle ( u_{x} - u_{x+r})^{3}\rangle$ (squares) that is the relevant expression in the structure function formulation of multifractality \cite{fp85}, it follows that surrogates and data should lie much closer to one another in the left hand column of Fig. \ref{fig.3}. The $z$-scores included in each panel show this is the case - there is no significant difference between data and surrogates for any case on the left hand side of the figure at the 5\% level, while all on the right are clearly significantly different at this level. Where difference are seen at the left-hand side, it is at the very smallest scales where there is inevitably some sensitivity to the random nature of the algorthm as differences are being taken over very small distances.

\section{Example applications}
There are a great range of potential applications of this algorithm as it provides a means for assessing confidence limits for any algorithm applied to multifractal data and, indeed, to techniques used to determine the multifractal properties of observed data. Here we consider two contrasting applications, both of which are of potential relevance to turbulence physics. 

\subsection{Oscillating singularities in time series}
Understanding the nature of any singularity or near-singularity (dissipation potentially smoothing discontinuities) in turbulence is an important area connecting topology to the Navier-Stokes equations \citep{moffat84,hunt91}. Related phenomena have been documented in studies of financial prices and rupture processes \citep{sornette02,sornette03}. We have already defined relevant cusp and chirp signals in terms of their pointwise H\"{o}lder, $\alpha_{p}(t)$, and local H\"{o}lder, $\alpha_{L}(t)$, exponents in (\ref{eq.chirp}), and described the coupled estimation of these exponents. In addition, note that the set of $\alpha_{p}(t)$ values is often represented by its singularity spectrum, $D(\alpha_{p})$ \citep{MUZY91}, but various multifractal spectra can be defined: the Haussdorff multifractal spectrum \citep{seuret06}; the Legendre transform spectrum \citep{MUZY91}; and, the large deviation spectrum \citep{aubry02}. Prominent additional modes in the latter, resulting in a loss of convexity, are indicative that the $\alpha_{L}$ are an important part of the signal. 

\subsubsection{Definition of artificial signals with a chirp}
Two test case signals are considered in this section: a fractional Brownian motion (fBm) with a chirp; and, a multifractional Brownian motion (mBm) with a chirp. They are defined for $0 \le t \le 1.2$ and consist of 24 000 discrete samples. The fBm has $\overline{\alpha}_{p} = 0.33$ (it is turbulence-like to the level of the second order structure function), while for the mBm we have $\alpha_{p}(t) =0.33 + 0.15 \sin 16 \pi t$. Both are discretised into 20 000 values over the support $0 \le t \le 1$. The chirp in both cases has a support of $-0.1 \le \chi \le 0.1$ discretised over 4000 values, and is given by $|\chi|^{\alpha_{+}} \sin \frac{1}{|\chi|^{\alpha_{\omega}}}$, with $\alpha_{\omega} = 1.4$ and $\alpha_{+} = 0.33$, using the notation in (\ref{eq.chirp}). Thus, $\alpha_{p} = 0.33$ and $\alpha_{L} \sim 0.14$. The chirp was inserted into the fBm or mBm near the center of the signal at a position that eliminated introducing any additional discontinuity (that the detection might be sensitive to). 

\begin{figure}[t]
\vspace*{2mm}
\begin{center}
\includegraphics[width=11.6cm]{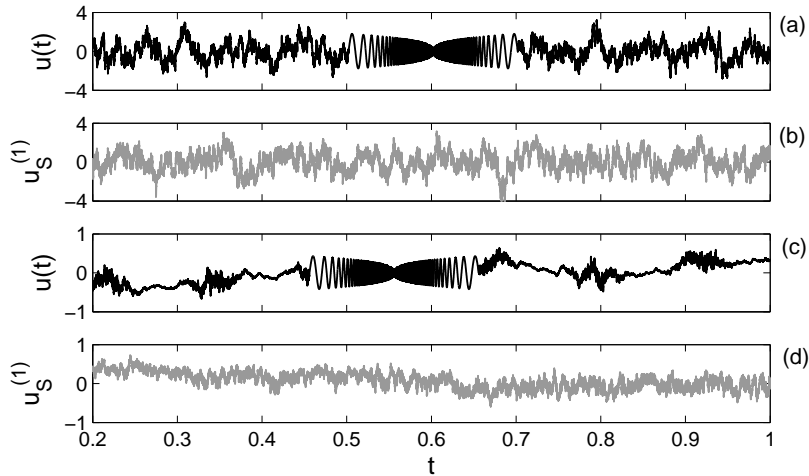}
\end{center}
\caption{A fractional Brownian motion with an embedded chirp is shown in black in panel (a). An example IAAFT surrogate is shown in gray in (b). Panel (c) shows a multifractional Brownian motion with an embedded chirp is shown in black. An IAAFT surrogate for this is shown in (d). The panels show a smaller segment of the full dataset in each case, which has a support of $0 \le t \le 1.2$.}
\label{fig.4}
\end{figure}

\begin{figure}[t]
\vspace*{2mm}
\begin{center}
\includegraphics[width=11.6cm]{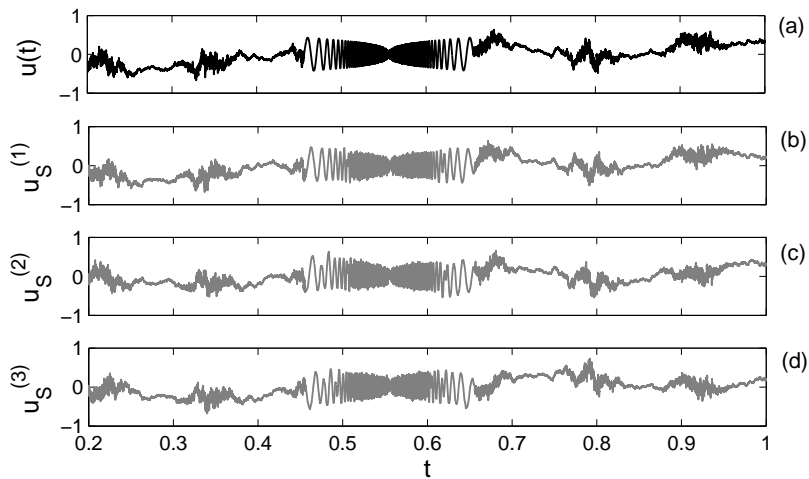}
\end{center}
\caption{A multifractional Brownian motion with an embedded chirp is shown in black in panel (a). Three example IAAWT surrogates are shown in gray in the lower panels. The preservation of the pointwise H\"{o}lder exponents and the location of the oscillating components is clear. The panels show a smaller segment of the full dataset in each case, which has a support of $0 \le t \le 1.2$.}
\label{fig.5}
\end{figure}

Figure \ref{fig.4} shows these signals in black focusing on the region of the chirp ($0.2 \le t \le 1$). IAAFT surrogates for these two signals are shown in gray in panels (b) and (d), and it is clear that the oscillating singularity and the intermittency have been destroyed. In contrast, Fig. \ref{fig.5} shows the signal from Fig. \ref{fig.4}c together with three IAAWT surrogates. It is clear that the IAAWT algorithm preserves the structure of the mBm, although the detailed behavior of the oscillation are lost (with only the broad characteristics giverned by $\alpha_{p}(t)$ preserved.  

\subsubsection{Oscillating singularity detection}
\begin{figure}[t]
\vspace*{2mm}
\begin{center}
\includegraphics[width=11.6cm]{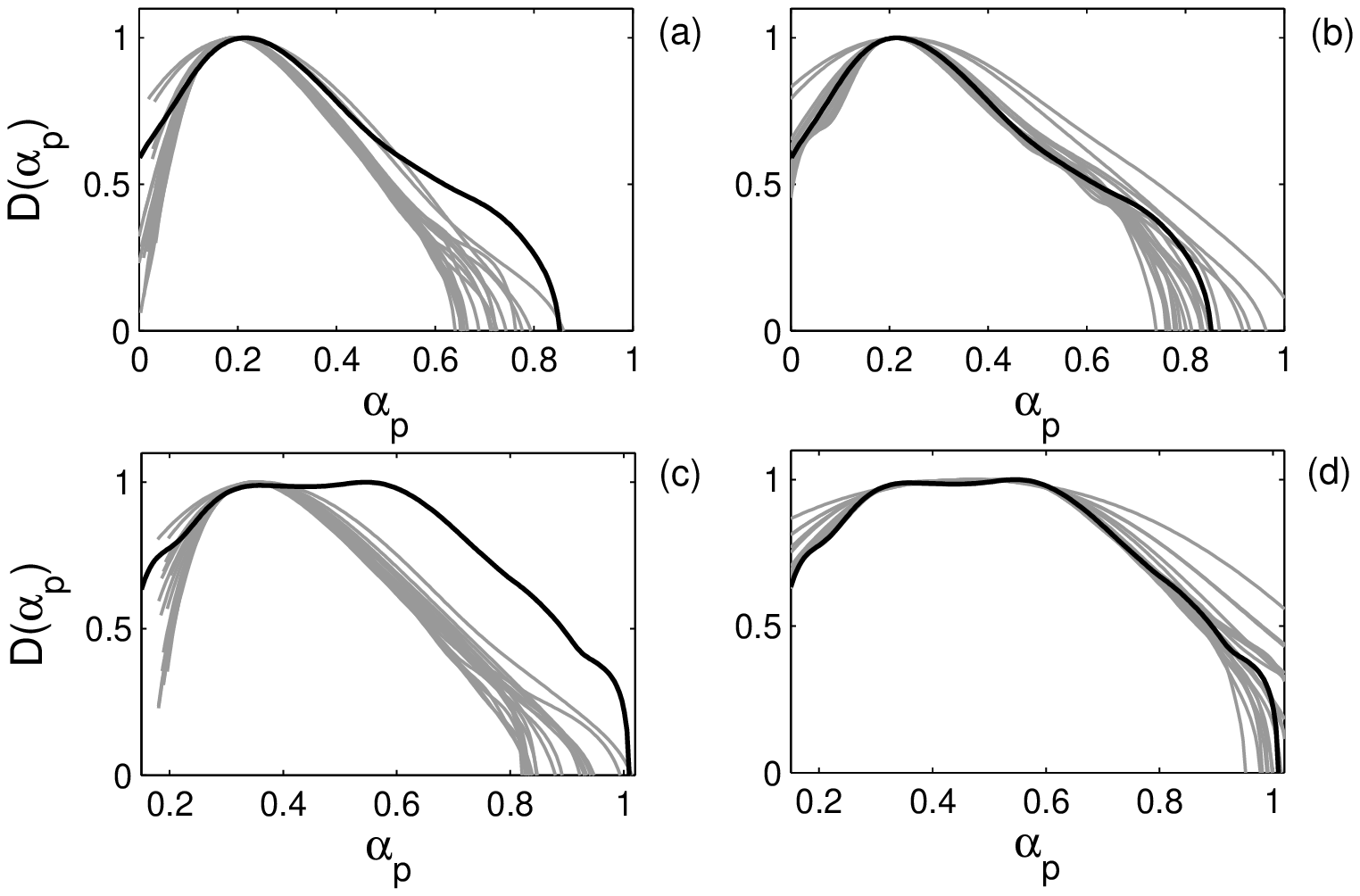}
\end{center}
\caption{Multifractal spectra, $D(\alpha_{p})$ estimated using the large deviation method (black) and compared to surrogate data (gray). Panels (a) and (b) are for a fractional Brownian motion with an embedded chirp, while (c) and (d) are for a multifractional Brownian motion with an embedded chirp. Panels (a) and (c) show IAAFT surrogate data, while IAAWT surrogates are used in (b) and (d). The multimodal structure of the spectrum indicates the presence of oscillating singularities in the original data in all cases, but this disappears in the IAAFT surrogates (although is retained by the IAAWT surrogates).}
\label{fig.6}
\end{figure}

In addition to the $K^{s,s^{'}}$ method discussed in the introduction and as noted above, the presence of oscillating singularities can be detected by a lack of convexity in the multifractal large deviation spectrum \citep{aubry02,seuret06}. Figure \ref{fig.6} shows the singularity spectra for the fBm + chirp (top row) and the mBm + chirp (bottom row) signals. In both cases there is a lack of convexity to the multifractal spectrum and in the latter case the secondary mode is as large as the primary mode centered at $\alpha_{p} \sim 0.33$. The left hand figures are compared to 19 IAAFT surrogates of the respective signals in gray, examples of which are given in Fig. 4. it is clear that convexity is recovered, indicating the removal of the oscillating singularity (the high values for $\alpha_{p}$ are a consequence of the edges of the chirp being smoother than $\alpha_{p} = 0.33$ in a pointwise sense). In contrast, the IAAWT surrogates in the right hand panels are clearly retaining the general $\alpha_{p}$ characteristics of the oscillation, with no significant difference existing between the $D(\alpha_{p}$ for the data and surrogates. 

Given that the general characteristics of the oscillation are retained in the IAAWT surrogates, we can use the $K^{s,s^{'}}$ frontiers (section IIB) to undertake a more detailed analysis and extract the oscillating singularity to be detected with statistical confidence. We estimated values for $\alpha_{p}(t)$ and $\alpha_{L}(t)$ for the mBm+chirp signal and for 19 IAAFT and IAAWT surrogates from the intersection of the $K^{s,s^{'}}$ frontier (estimated with 40 points) with $s^{'} = 0$ ($\alpha_{L}$) and with $s + s^{'} = 0$ ($\alpha_{p}$). This was undertaken for 1001 positions, centered on the middle of the chirp and traversing double the support of the chirp (i.e. every eighth value was calculated). Given a position, $t_{0}$, we may derive a \emph{z}-score estimate for the pointwise and local H\"{o}lder exponents at this point. Using $\alpha_{p}(t_{0})$ as an example, 
\begin{equation}
z_{\alpha_{p}}(t_{0}) = \frac{\alpha_{p}(t_{0}) - [\overline{\alpha}_{p}(t_{0})]_{S}}{\sigma[\alpha_{p}(t_{0})]_{S}}
\end{equation}
where the $S$ subscript indicates the operation is taken over the 19 values from the surrogate data, the overbar is the arithmetic mean and $\sigma$ is the standard deviation. Following \citet{DJ94}, an expected upper bound on these \emph{z}-scores (unit standard deviation by definition) is $(2 \log N / \sqrt{2})^{0.5}$.

\begin{figure}[t]
\vspace*{2mm}
\begin{center}
\includegraphics[width=11.6cm]{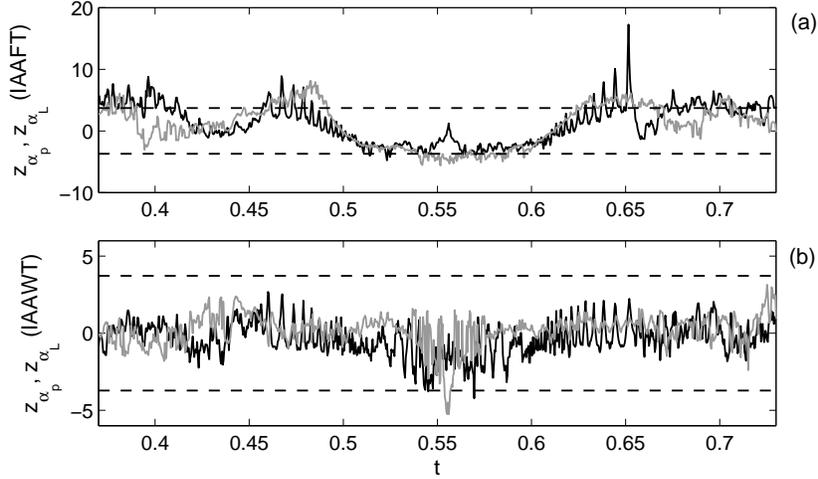}
\end{center}
\caption{Time series of \emph{z}-score estimates based on the the original data and the mean and standard deivation for surrogates for $\alpha_{p}$ (black) and $\alpha_{L}$ (gray). The upper panel is the results for IAAFT surrogates and the lower for IAAWT surrogates. An almost sure threshold \citep{DJ94} is indicated by the dashed lines.}
\label{fig.7}
\end{figure}

The results for $z_{\alpha_{p}}(t)$ (black) and $z_{\alpha_{L}}(t)$ (gray) are shown in Fig. \ref{fig.7}, using IAAFT (a) and IAAWT (b) surrogates. The expected upper bound on these values is shown by dashed lines on each panel. It is clear from Fig. \ref{fig.4}c that the center of the chirp lies at $t \sim 0.555$. Because IAAFT surrogates neither preserve multifractal structure or oscillating singularities, departures outside the expected upper bound arise at a variety of locations in Fig. \ref{fig.7}a for both $z_{\alpha_{p}}(t)$ and $z_{\alpha_{L}}(t)$. In contrast, apart from one datum at $z_{\alpha_{p}}(t \sim 0.57)$, the \emph{z}-scores for the pointwise H\"{o}lder exponents lie within the expected bound, which is to be expected for a multifractal preserving algorithm. In contrast, $z_{\alpha_{L}}$ remains within the bound, except at the center of the chirp, where a difference is observed across several consecutive values. Hence, the IAAWT algorithm preserves $\alpha_{p}(t)$ characteristics and, because it also fixes in place the position of particular features, locates a chirp-like feature in the correct place. However, it does not intrinsically preserve $\alpha_{L}$, meaning that the center of an oscillating singularity is extracted rather precisely. 

\subsection{Velocity-intermittency analyses for turbulence}
Kolmogorov's and Oboukhov's revised theories for turbulence, published in 1962 \citep{K62,Oboukhov62}, dealt with Landau's objection to the original 1941 theory \citep{K41} through the substitution of a Gaussian distribution for the dissipation with a log-normal distribution. This mathematical adjustment had the physical consequence of resulting in intermittency in statistical models of the turbulence cascade, which were interpreted as being due to the action of vortices in the flow \citep{frisch78}, and have been analysed from a multifractal perspective \citep{MS87,FV91,MUZY91}. However, another key change in the 1962 theory that has garnered less attention, is the explicit recognition of potential dependencies of the coefficients in the theory on the macroscale structure of the flow (as stated for equations 3 and 4 in \citet{K62}). Indeed, Frisch has argued that Kolmogorov was aware of such a potential dependence in 1941, but removed it from consideration at that time to derive a compact theory \citep{frisch05}. 

Given the potential importance of macroscale velocity to the cascade and the links between intermittency and coherent structures, potential velocity-intermittency coupling in experimentally measured flows may be analyzed using a method built around the notion of quadrants,which are typically used to describe the structure of boundary-layers \citep{nezu77,bt86}. However, in the velocity-intermittency approach, the quadrants are based on the joint distribution function for the longitudinal velocity component, $u$, and its pointwise H\"{o}lder regularity, $\alpha_{p}(u)$, rather than for $u$ and the vertical velocity component, $v$ \citep{k12}. This method has had some success in classifying different canonical flows, demonstrating how turbulent flows forced in different ways have different characteristics \citep{k13}, with results interpretable in terms of the behavior of coherent structures \citep{k16}. 

The velocity-intermittency quadrants are defined and analyzed as follows:
\begin{enumerate}
\item For a given velocity time series, $u(t)$, calculate the corresponding $\alpha_{p}(u)$ and scale these two terms by their respective means and standard deviations, i.e. $u^{*} = (u - \overline{u})/\sigma(u)$;
\item Define threshold`hole-sizes', $0 \le H \le H_{\mbox{max}}$, where $H_{\mbox{max}}$ is the largest value justifiable based on the number of values, in the time series, $N$, and $H_{\mbox{max}} \propto \mbox{log} N$, and a threshold exceedance arises when $|u^{*}\alpha_{p}(u)^{*}| > H \sigma(u)\sigma(\alpha_{p}(u))$;
\item For each $H$, form the probability of quadrant occupancy, $p_{Q}$, which is the number of $u^{*}-\alpha_{p}(u)^{*}$ pairs exceeding $H$ in a given quadrant, $Q$, divided by the number of pairs exceeding this $H$ in all four quadrants;
\item Approximate the relation between $p_{Q}$ and $H$ by its slope, $d p_{Q}/ d H$ and characterize a particular turbulent flow by its four values for $d p_{Q}/ d H$.
\end{enumerate} 
Given a particular set of measurements for $u(t)$ as a function of time, then the distribution function of $u(t)$ will have some impact on the joint distribution for $u$ and $\alpha_{p}(u)$. For example, if $u(t)$ is positively skewed, then quadrants Q1 ($u'>0, \alpha_{p}(u)' > 0$) and Q4 ($u'>0, \alpha_{p}(u)' < 0$) are liable to dominate the extremal statistics. Hence, there is value in placing confidence limits on the derived values for $d p_{Q}/ d H$ and IAAWT surrogates provide a means to do this as the preservation of the values for $u(t)$ and $\alpha_{p}(u)$ independently, means one can determine if the $d p_{Q}/ d H$ could have arisen by chance given the observed $u(t)^{*}$ and $\alpha_{p}(u)^{*}$. (Note that this is a test for velocity-intermittency given the velocity and intermittency. As the nature of the forcing will impact on the structure for $\alpha_{p}(u)^{*}$, in particular, the observed $d p_{Q}/ d H$ may still be highly informative and significant physically, even if their joint structure is not significant when conditioned on values for the marginals).

\subsubsection{Analysis of a turbulent wake}
The data analysed here were measured in a 1 m cross-section wind tunnel with a hot wire anenometer at 5 kHz in the wake of a $h = 100$ mm high fence, immersed in a neutral boundary-layer with a thickness of $\sim 200 mm$. The fence consisted of a 10 mm bottom gap, followed by nine 5.3 mm horizontal elements, with a 5.3 mm space between each. Hence, the center of the fence, proper is 55 mm above the wall ($y/h = 0.55$), and we recorded data for several hundred integral scales at distances downwind of the fence, $x$, ranging from $1.25 \le x/h \le 10$. Further details on the wind tunnel, the experiments and properties of the wake structure are given by \citet{k12b}. 

\begin{figure}[t]
\vspace*{2mm}
\begin{center}
\includegraphics[width=11.6cm]{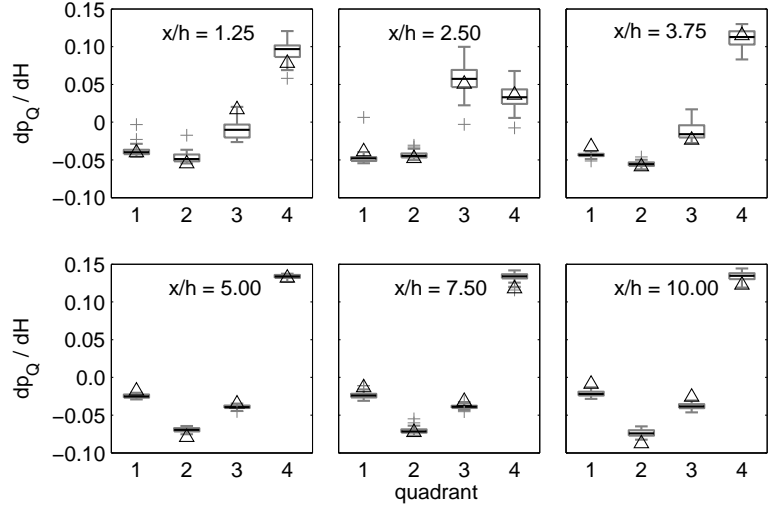}
\end{center}
\caption{Boxplots of the results for the velocity-intermittency quadrant analysis at various dimensionless distances downwind of the fence. Values for the actual data are shown as hollow triangles. The boxes themselves are based on the values for 39 IAAWT surrogates, with outliers indicated by a +, the box consisting of the median (center line) and first and third quartiles (bottom and top edges of the box, respectively), and the whiskers extending up to 1.5 times the interquartile deviation.}
\label{fig.boxp}
\end{figure}

\begin{figure}
\noindent\includegraphics[width=8.6cm]{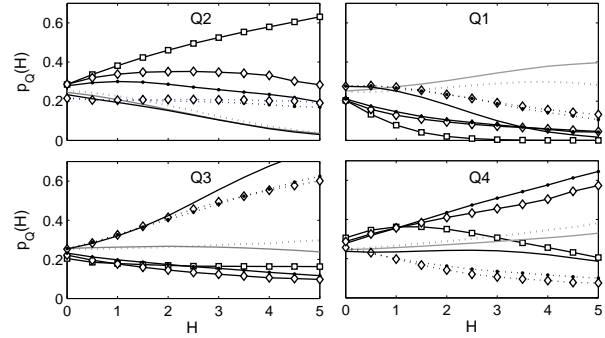}
\caption{An analysis of velocity-intermittency from various experiments. The various lines correspond to data from a turbulent jet experiment \citep{renner01} (solid black line with hollow squares), wake data at 8.5 ms$^{-1}$ (grey dotted) and 24.3 ms$^{-1}$ (grey) \citep{stresing10}, and data below 150 wall units (solid black lines) and above 700 wall units (dotted black lines) at 6 ms$^{-1}$ (hollow diamonds) and 8 ms$^{-1}$ (solid circles) for a zero pressure gradient boundary-layer \citep{k12}. Data for flow over bed-forms \citep{k13} are shown as a solid black line. This figure is modified from: Keylock, C.J., Singh, A., Foufoula-Georgiou, E. 2013. The influence of bedforms on the velocity-intermittency structure of turbulent flow over a gravel bed, Geophysical Research Letters 40, 1-5, doi:10.1002/grl.50337. (copyright American Geophysical Union) and is reproduced with the permission of the AGU.}
\label{fig.GRL}
\end{figure}

Results for the velocity-intermittency structure for these data at various $x/h$ and $y/h=0.55$ are shown in Fig. \ref{fig.boxp} as hollow triangles. The dominance of positive slopes for quadrant 4 ($u^{*}>0,\alpha_{p}(u)^{*}<0$) is clear, meaning that the extremal events are of high relative velocity and intermittency. This may be contrasted with results for other flows shown in Fig. \ref{fig.GRL}. For example, while the far field wake generated by flow about a cylinder (from the experiment described in \citet{stresing10}) also shows a positive slope for quadrant 4 (the grey, dotted and grey lines), the value for $d p_{Q}/dH \sim 0.03$ is much smaller. In fact, the greater positive slope for these data is in quadrant 1. In Fig. \ref{fig.GRL} it is the boundary-layer data near the wall that shows the strongest positive slope in quadrant 4, with $dp_{Q}/dH \sim 0.07$. Given that the plane of shear for the fence wake is oriented in the longitudinal-vertical (i.e. similar to the Reynolds stress gradient in a boundary-layer) rather than the longitudinal-transverse plane for the cylinder wake, and that the Reynolds stresses in this wake region are much greater than for the boundary-layer case, this quadrant 4 dominance for the fence wake is to be expected. 

\begin{figure}[t]
\vspace*{2mm}
\begin{center}
\includegraphics[width=11.6cm]{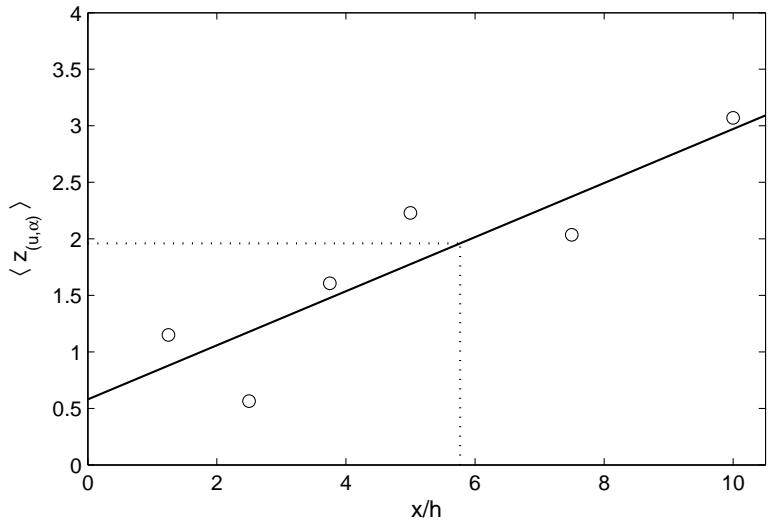}
\end{center}
\caption{A summary of the results in Fig. \ref{fig.boxp} where the distance between the data and the mean of the surrogates, normalized by the standard deviation of the surrogates (hence, a $z$-score, $z_{(u,\alpha)}$) is reported, with an average taken over all four quadrants.}
\label{fig.zsc}
\end{figure}

In addition to the results for the wind tunnel data in Fig. \ref{fig.boxp}, we also show those obtained by generating 39 IAAWT surrogates for $u$ at the six different positions. These results are summarized in boxplot form and it is clear that the surrogate data also exhibit the pattern of quadrant 4 dominance. Thus, the prescribed distribution function for $u$ and the prescribed H\"{o}lder function constrains the data to a quadrant four dominating response. However, the extent to which the surrogates can select different values for $dp_{Q}/dH$ is much more constrained for $x/h > 5$ than closer to the fence. If $z_{(u,\alpha)}(Q)$ is the $z$-score for the $dp_{Q}/dH$ for a given quadrant, then $\langle z_{(u,\alpha)} \rangle$ is the average $z$-score over all quadrants at a given $x/h$. This provides a summary measure of the extent to which the data and surrogates are different and for $x/h \le 10$, it is clear from Fig. \ref{fig.zsc} that this value grows downwind of the fence, with $\langle z_{(u,\alpha)} \rangle > 1.96$ at $x/h \gtrsim 6$. This is despite any major increase in $dp_{Q=4}/dH$ over this range of $x/h$, indicating that the velocity-intermittency structure continues to evolve relative to appropriately constrained  surrogates even though the actual structure of the four values for $dp_{Q}/dH$ is relatively invariant for $x/h > 5$.

\section{Conclusion}
We have presented a surrogate data algorithm designed to preserve, not only the multifractal characteristics of a time-series or image (the probability function for the pointwise H\"{o}lder regularity, $\alpha_{p}$), but the observed distribution of values (in time or space). Compared to an existing algorithm \citep{PALUS} ours replicates the distribution function for the H\"{o}lder exponents more accurately (Fig. \ref{fig.2}) while also preserving the distribution function of the raw data values, making our algorithm more akin to the well-known IAAFT algorithm \citep{ss96}.

Hence, not only can this algorithm be used to demonstrate bootstrapped data for multifractal processes, it may be used for analysis of processes where the outputs contain greater structure than a random multifractal process \citep{aubry02}. We have presented two such examples: the detection of oscillating singularities in a time series; and, the analysis of turbulence data that exhibit velocity and intermittency coupling. The latter example is particularly exciting in the context of recent extensions of fractional Brownian and multifractional Brownian motions to \emph{self-regulating processes} where as, with our turbulence analysis, local regularity is a function of amplitude \citep{levyvehel13,echelard15}. Our work complements these studies, which establish the mathematical basis for the formulation and estimation of such processes, by providing a means to test if any observed coupling between regularity and amplitude is significant. Thus, it provides a practical criterion for assessing if the modelling of biophysical time-series, planetary topography or turbulence processes with complex forcings can be enhanced by modeling them as self-regulating processes.

\acknowledgments
This work was supported by NERC grant NE/F00415X/1, EPSRC grant EP/K007688/1 and Royal Academy of Engineering/Leverhulme Senior Research Fellowship  LTSRF1516-12-89.

\bibliography{PRE_2016}

\begin{thebibliography}{75}%
\makeatletter
\providecommand \@ifxundefined [1]{%
 \@ifx{#1\undefined}
}%
\providecommand \@ifnum [1]{%
 \ifnum #1\expandafter \@firstoftwo
 \else \expandafter \@secondoftwo
 \fi
}%
\providecommand \@ifx [1]{%
 \ifx #1\expandafter \@firstoftwo
 \else \expandafter \@secondoftwo
 \fi
}%
\providecommand \natexlab [1]{#1}%
\providecommand \enquote  [1]{``#1''}%
\providecommand \bibnamefont  [1]{#1}%
\providecommand \bibfnamefont [1]{#1}%
\providecommand \citenamefont [1]{#1}%
\providecommand \href@noop [0]{\@secondoftwo}%
\providecommand \href [0]{\begingroup \@sanitize@url \@href}%
\providecommand \@href[1]{\@@startlink{#1}\@@href}%
\providecommand \@@href[1]{\endgroup#1\@@endlink}%
\providecommand \@sanitize@url [0]{\catcode `\\12\catcode `\$12\catcode
  `\&12\catcode `\#12\catcode `\^12\catcode `\_12\catcode `\%12\relax}%
\providecommand \@@startlink[1]{}%
\providecommand \@@endlink[0]{}%
\providecommand \url  [0]{\begingroup\@sanitize@url \@url }%
\providecommand \@url [1]{\endgroup\@href {#1}{\urlprefix }}%
\providecommand \urlprefix  [0]{URL }%
\providecommand \Eprint [0]{\href }%
\providecommand \doibase [0]{http://dx.doi.org/}%
\providecommand \selectlanguage [0]{\@gobble}%
\providecommand \bibinfo  [0]{\@secondoftwo}%
\providecommand \bibfield  [0]{\@secondoftwo}%
\providecommand \translation [1]{[#1]}%
\providecommand \BibitemOpen [0]{}%
\providecommand \bibitemStop [0]{}%
\providecommand \bibitemNoStop [0]{.\EOS\space}%
\providecommand \EOS [0]{\spacefactor3000\relax}%
\providecommand \BibitemShut  [1]{\csname bibitem#1\endcsname}%
\let\auto@bib@innerbib\@empty
\bibitem [{\citenamefont {{Theiler}}\ \emph {et~al.}(1992)\citenamefont
  {{Theiler}}, \citenamefont {{Eubank}}, \citenamefont {{Longtin}},
  \citenamefont {{Galdrikian}},\ and\ \citenamefont {{Farmer}}}]{T92}%
  \BibitemOpen
  \bibfield  {author} {\bibinfo {author} {\bibfnamefont {J.}~\bibnamefont
  {{Theiler}}}, \bibinfo {author} {\bibfnamefont {S.}~\bibnamefont {{Eubank}}},
  \bibinfo {author} {\bibfnamefont {A.}~\bibnamefont {{Longtin}}}, \bibinfo
  {author} {\bibfnamefont {B.}~\bibnamefont {{Galdrikian}}}, \ and\ \bibinfo
  {author} {\bibfnamefont {J.}~\bibnamefont {{Farmer}}},\ }\href@noop {}
  {\bibfield  {journal} {\bibinfo  {journal} {Physica D}\ }\textbf {\bibinfo
  {volume} {58}},\ \bibinfo {pages} {77} (\bibinfo {year} {1992})}\BibitemShut
  {NoStop}%
\bibitem [{\citenamefont {{Prichard}}\ and\ \citenamefont
  {{Theiler}}(1994)}]{PT94}%
  \BibitemOpen
  \bibfield  {author} {\bibinfo {author} {\bibfnamefont {D.}~\bibnamefont
  {{Prichard}}}\ and\ \bibinfo {author} {\bibfnamefont {J.}~\bibnamefont
  {{Theiler}}},\ }\href@noop {} {\bibfield  {journal} {\bibinfo  {journal}
  {Phys. Rev. Lett.}\ }\textbf {\bibinfo {volume} {73}},\ \bibinfo {pages}
  {951} (\bibinfo {year} {1994})}\BibitemShut {NoStop}%
\bibitem [{\citenamefont {{Schreiber}}\ and\ \citenamefont
  {{Schmitz}}(1996)}]{ss96}%
  \BibitemOpen
  \bibfield  {author} {\bibinfo {author} {\bibfnamefont {T.}~\bibnamefont
  {{Schreiber}}}\ and\ \bibinfo {author} {\bibfnamefont {A.}~\bibnamefont
  {{Schmitz}}},\ }\href@noop {} {\bibfield  {journal} {\bibinfo  {journal}
  {Phys. Rev. Lett.}\ }\textbf {\bibinfo {volume} {77}},\ \bibinfo {pages}
  {635} (\bibinfo {year} {1996})}\BibitemShut {NoStop}%
\bibitem [{\citenamefont {{Small}}\ \emph {et~al.}(2001)\citenamefont
  {{Small}}, \citenamefont {{Yu}},\ and\ \citenamefont {{Harrison}}}]{small01}%
  \BibitemOpen
  \bibfield  {author} {\bibinfo {author} {\bibfnamefont {M.}~\bibnamefont
  {{Small}}}, \bibinfo {author} {\bibfnamefont {D.}~\bibnamefont {{Yu}}}, \
  and\ \bibinfo {author} {\bibfnamefont {R.}~\bibnamefont {{Harrison}}},\
  }\href@noop {} {\bibfield  {journal} {\bibinfo  {journal} {Phys. Rev. Lett.}\
  }\textbf {\bibinfo {volume} {87}},\ \bibinfo {pages} {188101} (\bibinfo
  {year} {2001})}\BibitemShut {NoStop}%
\bibitem [{\citenamefont {{Keylock}}(2006)}]{k06}%
  \BibitemOpen
  \bibfield  {author} {\bibinfo {author} {\bibfnamefont {C.~J.}\ \bibnamefont
  {{Keylock}}},\ }\href@noop {} {\bibfield  {journal} {\bibinfo  {journal}
  {Phys. Rev. E}\ }\textbf {\bibinfo {volume} {73}},\ \bibinfo {pages} {036707}
  (\bibinfo {year} {2006})}\BibitemShut {NoStop}%
\bibitem [{\citenamefont {{Theiler}}\ and\ \citenamefont
  {{Rapp}}(1996)}]{theiler96}%
  \BibitemOpen
  \bibfield  {author} {\bibinfo {author} {\bibfnamefont {J.}~\bibnamefont
  {{Theiler}}}\ and\ \bibinfo {author} {\bibfnamefont {P.~E.}\ \bibnamefont
  {{Rapp}}},\ }\href@noop {} {\bibfield  {journal} {\bibinfo  {journal}
  {Electroencephal. Clin. Neurophys.}\ }\textbf {\bibinfo {volume} {98}},\
  \bibinfo {pages} {213} (\bibinfo {year} {1996})}\BibitemShut {NoStop}%
\bibitem [{\citenamefont {{Govindan}}\ \emph {et~al.}(1998)\citenamefont
  {{Govindan}}, \citenamefont {{Narayanan}},\ and\ \citenamefont
  {{Gopinathan}}}]{govindan98}%
  \BibitemOpen
  \bibfield  {author} {\bibinfo {author} {\bibfnamefont {R.~B.}\ \bibnamefont
  {{Govindan}}}, \bibinfo {author} {\bibfnamefont {K.}~\bibnamefont
  {{Narayanan}}}, \ and\ \bibinfo {author} {\bibfnamefont {M.~S.}\ \bibnamefont
  {{Gopinathan}}},\ }\href@noop {} {\bibfield  {journal} {\bibinfo  {journal}
  {Chaos}\ }\textbf {\bibinfo {volume} {8}},\ \bibinfo {pages} {495} (\bibinfo
  {year} {1998})}\BibitemShut {NoStop}%
\bibitem [{\citenamefont {{Ivanov}}\ \emph {et~al.}(1999)\citenamefont
  {{Ivanov}}, \citenamefont {{Amaral}}, \citenamefont {{Goldberger}},
  \citenamefont {{Havlin}}, \citenamefont {{Rosenblum}}, \citenamefont
  {{Struzik}},\ and\ \citenamefont {{Stanley}}}]{ivanov99}%
  \BibitemOpen
  \bibfield  {author} {\bibinfo {author} {\bibfnamefont {P.~C.}\ \bibnamefont
  {{Ivanov}}}, \bibinfo {author} {\bibfnamefont {L.~A.~N.}\ \bibnamefont
  {{Amaral}}}, \bibinfo {author} {\bibfnamefont {A.~L.}\ \bibnamefont
  {{Goldberger}}}, \bibinfo {author} {\bibfnamefont {S.}~\bibnamefont
  {{Havlin}}}, \bibinfo {author} {\bibfnamefont {M.~G.}\ \bibnamefont
  {{Rosenblum}}}, \bibinfo {author} {\bibfnamefont {Z.~R.}\ \bibnamefont
  {{Struzik}}}, \ and\ \bibinfo {author} {\bibfnamefont {H.~E.}\ \bibnamefont
  {{Stanley}}},\ }\href {\doibase 10.1038/20924} {\bibfield  {journal}
  {\bibinfo  {journal} {Nature}\ }\textbf {\bibinfo {volume} {399}},\ \bibinfo
  {pages} {461} (\bibinfo {year} {1999})}\BibitemShut {NoStop}%
\bibitem [{\citenamefont {{Basu}}\ \emph {et~al.}(2004)\citenamefont {{Basu}},
  \citenamefont {{Foufoula-Georgiou}},\ and\ \citenamefont
  {{Port\'{e}-Agel}}}]{basu04}%
  \BibitemOpen
  \bibfield  {author} {\bibinfo {author} {\bibfnamefont {S.}~\bibnamefont
  {{Basu}}}, \bibinfo {author} {\bibfnamefont {E.}~\bibnamefont
  {{Foufoula-Georgiou}}}, \ and\ \bibinfo {author} {\bibfnamefont
  {F.}~\bibnamefont {{Port\'{e}-Agel}}},\ }\href@noop {} {\bibfield  {journal}
  {\bibinfo  {journal} {Phys. Rev. E}\ }\textbf {\bibinfo {volume} {70}}
  (\bibinfo {year} {2004})}\BibitemShut {NoStop}%
\bibitem [{\citenamefont {{Poggi}}\ \emph {et~al.}(2004)\citenamefont
  {{Poggi}}, \citenamefont {{Porporato}}, \citenamefont {{Ridolfi}},
  \citenamefont {{Albertson}},\ and\ \citenamefont {{Katul}}}]{poggi04}%
  \BibitemOpen
  \bibfield  {author} {\bibinfo {author} {\bibfnamefont {D.}~\bibnamefont
  {{Poggi}}}, \bibinfo {author} {\bibfnamefont {A.}~\bibnamefont
  {{Porporato}}}, \bibinfo {author} {\bibfnamefont {L.}~\bibnamefont
  {{Ridolfi}}}, \bibinfo {author} {\bibfnamefont {J.~D.}\ \bibnamefont
  {{Albertson}}}, \ and\ \bibinfo {author} {\bibfnamefont {G.~G.}\ \bibnamefont
  {{Katul}}},\ }\href {\doibase 10.1029/2003GL018611} {\bibfield  {journal}
  {\bibinfo  {journal} {Geophys. Res. Lett.}\ }\textbf {\bibinfo {volume}
  {31}},\ \bibinfo {pages} {L05102} (\bibinfo {year} {2004})}\BibitemShut
  {NoStop}%
\bibitem [{\citenamefont {{Volobuev}}\ and\ \citenamefont
  {{Makarenko}}(2008)}]{volobuev08}%
  \BibitemOpen
  \bibfield  {author} {\bibinfo {author} {\bibfnamefont {D.~M.}\ \bibnamefont
  {{Volobuev}}}\ and\ \bibinfo {author} {\bibfnamefont {N.~G.}\ \bibnamefont
  {{Makarenko}}},\ }\href {\doibase 10.1007/s11207-008-9167-y} {\bibfield
  {journal} {\bibinfo  {journal} {Solar Physics}\ }\textbf {\bibinfo {volume}
  {249}},\ \bibinfo {pages} {121} (\bibinfo {year} {2008})}\BibitemShut
  {NoStop}%
\bibitem [{\citenamefont {{Keylock}}\ \emph
  {et~al.}(2014{\natexlab{a}})\citenamefont {{Keylock}}, \citenamefont
  {{Lane}},\ and\ \citenamefont {{Richards}}}]{k14b}%
  \BibitemOpen
  \bibfield  {author} {\bibinfo {author} {\bibfnamefont {C.~J.}\ \bibnamefont
  {{Keylock}}}, \bibinfo {author} {\bibfnamefont {S.~N.}\ \bibnamefont
  {{Lane}}}, \ and\ \bibinfo {author} {\bibfnamefont {K.~S.}\ \bibnamefont
  {{Richards}}},\ }\href {\doibase 10.1002/2012JF002698} {\bibfield  {journal}
  {\bibinfo  {journal} {J. Geophys. Res.}\ }\textbf {\bibinfo {volume} {119}},\
  \bibinfo {pages} {264} (\bibinfo {year} {2014}{\natexlab{a}})}\BibitemShut
  {NoStop}%
\bibitem [{\citenamefont {{Schreiber}}\ and\ \citenamefont
  {{Schmitz}}(1997)}]{ss97}%
  \BibitemOpen
  \bibfield  {author} {\bibinfo {author} {\bibfnamefont {T.}~\bibnamefont
  {{Schreiber}}}\ and\ \bibinfo {author} {\bibfnamefont {A.}~\bibnamefont
  {{Schmitz}}},\ }\href@noop {} {\bibfield  {journal} {\bibinfo  {journal}
  {Phys. Rev. E}\ }\textbf {\bibinfo {volume} {55}},\ \bibinfo {pages} {5443}
  (\bibinfo {year} {1997})}\BibitemShut {NoStop}%
\bibitem [{\citenamefont {{Keylock}}(2010)}]{k10}%
  \BibitemOpen
  \bibfield  {author} {\bibinfo {author} {\bibfnamefont {C.~J.}\ \bibnamefont
  {{Keylock}}},\ }\href@noop {} {\bibfield  {journal} {\bibinfo  {journal}
  {Nonlinear Proc.Geophys.}\ }\textbf {\bibinfo {volume} {17}},\ \bibinfo
  {pages} {615} (\bibinfo {year} {2010})}\BibitemShut {NoStop}%
\bibitem [{\citenamefont {{Breakspear}}\ \emph {et~al.}(2003)\citenamefont
  {{Breakspear}}, \citenamefont {M.{Brammer}},\ and\ \citenamefont
  {{Robinson}}}]{B03}%
  \BibitemOpen
  \bibfield  {author} {\bibinfo {author} {\bibfnamefont {M.}~\bibnamefont
  {{Breakspear}}}, \bibinfo {author} {\bibnamefont {M.{Brammer}}}, \ and\
  \bibinfo {author} {\bibfnamefont {P.}~\bibnamefont {{Robinson}}},\
  }\href@noop {} {\bibfield  {journal} {\bibinfo  {journal} {Physica D}\
  }\textbf {\bibinfo {volume} {182}},\ \bibinfo {pages} {1} (\bibinfo {year}
  {2003})}\BibitemShut {NoStop}%
\bibitem [{\citenamefont {{Keylock}}(2007)}]{k07}%
  \BibitemOpen
  \bibfield  {author} {\bibinfo {author} {\bibfnamefont {C.~J.}\ \bibnamefont
  {{Keylock}}},\ }\href@noop {} {\bibfield  {journal} {\bibinfo  {journal}
  {Physica D}\ }\textbf {\bibinfo {volume} {225}},\ \bibinfo {pages} {219}
  (\bibinfo {year} {2007})}\BibitemShut {NoStop}%
\bibitem [{\citenamefont {{Borgnat}}\ and\ \citenamefont
  {{Flandrin}}(2009)}]{flandrin09}%
  \BibitemOpen
  \bibfield  {author} {\bibinfo {author} {\bibfnamefont {P.}~\bibnamefont
  {{Borgnat}}}\ and\ \bibinfo {author} {\bibfnamefont {P.}~\bibnamefont
  {{Flandrin}}},\ }\href {\doibase doi:10.1088/1742-5468/2009/01/P01001}
  {\bibfield  {journal} {\bibinfo  {journal} {J. Stat. Mech.}\ } (\bibinfo
  {year} {2009}),\ doi:10.1088/1742-5468/2009/01/P01001}\BibitemShut {NoStop}%
\bibitem [{\citenamefont {{Keylock}}\ \emph {et~al.}(2011)\citenamefont
  {{Keylock}}, \citenamefont {{Tokyay}},\ and\ \citenamefont
  {{Constantinescu}}}]{k11}%
  \BibitemOpen
  \bibfield  {author} {\bibinfo {author} {\bibfnamefont {C.~J.}\ \bibnamefont
  {{Keylock}}}, \bibinfo {author} {\bibfnamefont {T.~E.}\ \bibnamefont
  {{Tokyay}}}, \ and\ \bibinfo {author} {\bibfnamefont {G.}~\bibnamefont
  {{Constantinescu}}},\ }\href {\doibase 10.1080/14685248.2011.636047}
  {\bibfield  {journal} {\bibinfo  {journal} {J. Turbul.}\ }\textbf {\bibinfo
  {volume} {12}},\ \bibinfo {pages} {N45} (\bibinfo {year} {2011})}\BibitemShut
  {NoStop}%
\bibitem [{\citenamefont {{Keylock}}(2012)}]{k12c}%
  \BibitemOpen
  \bibfield  {author} {\bibinfo {author} {\bibfnamefont {C.~J.}\ \bibnamefont
  {{Keylock}}},\ }\href {\doibase 10.1029/2012WR011923} {\bibfield  {journal}
  {\bibinfo  {journal} {Water Resour. Res.}\ }\textbf {\bibinfo {volume} {48}}
  (\bibinfo {year} {2012}),\ 10.1029/2012WR011923}\BibitemShut {NoStop}%
\bibitem [{\citenamefont {{Keylock}}\ \emph
  {et~al.}(2014{\natexlab{b}})\citenamefont {{Keylock}}, \citenamefont
  {{Singh}},\ and\ \citenamefont {{Foufoula-Georgiou}}}]{k14}%
  \BibitemOpen
  \bibfield  {author} {\bibinfo {author} {\bibfnamefont {C.~J.}\ \bibnamefont
  {{Keylock}}}, \bibinfo {author} {\bibfnamefont {A.}~\bibnamefont {{Singh}}},
  \ and\ \bibinfo {author} {\bibfnamefont {E.}~\bibnamefont
  {{Foufoula-Georgiou}}},\ }\href {\doibase 10.1002/2013JF002999} {\bibfield
  {journal} {\bibinfo  {journal} {J. Geophys. Res.}\ }\textbf {\bibinfo
  {volume} {119}},\ \bibinfo {pages} {682} (\bibinfo {year}
  {2014}{\natexlab{b}})}\BibitemShut {NoStop}%
\bibitem [{\citenamefont {{Palu\u{s}}}(2008)}]{PALUS}%
  \BibitemOpen
  \bibfield  {author} {\bibinfo {author} {\bibfnamefont {M.}~\bibnamefont
  {{Palu\u{s}}}},\ }\href@noop {} {\bibfield  {journal} {\bibinfo  {journal}
  {Phys. Rev. Lett.}\ }\textbf {\bibinfo {volume} {101}},\ \bibinfo {pages}
  {134101} (\bibinfo {year} {2008})}\BibitemShut {NoStop}%
\bibitem [{\citenamefont {{Gagnon}}\ \emph {et~al.}(2003)\citenamefont
  {{Gagnon}}, \citenamefont {{Lovejoy}},\ and\ \citenamefont
  {{Schertzer}}}]{GLS03}%
  \BibitemOpen
  \bibfield  {author} {\bibinfo {author} {\bibfnamefont {J.}~\bibnamefont
  {{Gagnon}}}, \bibinfo {author} {\bibfnamefont {S.}~\bibnamefont {{Lovejoy}}},
  \ and\ \bibinfo {author} {\bibfnamefont {D.}~\bibnamefont {{Schertzer}}},\
  }\href@noop {} {\bibfield  {journal} {\bibinfo  {journal} {Europhys. Lett.}\
  }\textbf {\bibinfo {volume} {62}},\ \bibinfo {pages} {801} (\bibinfo {year}
  {2003})}\BibitemShut {NoStop}%
\bibitem [{\citenamefont {{Venugopal}}\ \emph {et~al.}(2006)\citenamefont
  {{Venugopal}}, \citenamefont {{Roux}}, \citenamefont {{Foufoula-Georgiou}},\
  and\ \citenamefont {{Arn\'{e}odo}}}]{ven06}%
  \BibitemOpen
  \bibfield  {author} {\bibinfo {author} {\bibfnamefont {V.}~\bibnamefont
  {{Venugopal}}}, \bibinfo {author} {\bibfnamefont {S.}~\bibnamefont {{Roux}}},
  \bibinfo {author} {\bibfnamefont {E.}~\bibnamefont {{Foufoula-Georgiou}}}, \
  and\ \bibinfo {author} {\bibfnamefont {A.}~\bibnamefont {{Arn\'{e}odo}}},\
  }\href@noop {} {\bibfield  {journal} {\bibinfo  {journal} {Phys. Lett. A}\
  }\textbf {\bibinfo {volume} {348}},\ \bibinfo {pages} {335} (\bibinfo {year}
  {2006})}\BibitemShut {NoStop}%
\bibitem [{\citenamefont {{Matia}}\ \emph {et~al.}(2003)\citenamefont
  {{Matia}}, \citenamefont {{Ashkenazy}},\ and\ \citenamefont
  {{Stanley}}}]{MATIA}%
  \BibitemOpen
  \bibfield  {author} {\bibinfo {author} {\bibfnamefont {K.}~\bibnamefont
  {{Matia}}}, \bibinfo {author} {\bibfnamefont {Y.}~\bibnamefont
  {{Ashkenazy}}}, \ and\ \bibinfo {author} {\bibfnamefont {H.}~\bibnamefont
  {{Stanley}}},\ }\href@noop {} {\bibfield  {journal} {\bibinfo  {journal}
  {Europhys. Lett.}\ }\textbf {\bibinfo {volume} {61}},\ \bibinfo {pages} {422}
  (\bibinfo {year} {2003})}\BibitemShut {NoStop}%
\bibitem [{\citenamefont {{Klesse}}\ and\ \citenamefont
  {{Metzler}}(1995)}]{KM95}%
  \BibitemOpen
  \bibfield  {author} {\bibinfo {author} {\bibfnamefont {R.}~\bibnamefont
  {{Klesse}}}\ and\ \bibinfo {author} {\bibfnamefont {M.}~\bibnamefont
  {{Metzler}}},\ }\href@noop {} {\bibfield  {journal} {\bibinfo  {journal}
  {Europhys. Lett.}\ }\textbf {\bibinfo {volume} {32}},\ \bibinfo {pages} {229}
  (\bibinfo {year} {1995})}\BibitemShut {NoStop}%
\bibitem [{\citenamefont {{Jaffard}}\ and\ \citenamefont
  {{Meyer}}(1996)}]{jaffard96}%
  \BibitemOpen
  \bibfield  {author} {\bibinfo {author} {\bibfnamefont {S.}~\bibnamefont
  {{Jaffard}}}\ and\ \bibinfo {author} {\bibfnamefont {Y.}~\bibnamefont
  {{Meyer}}},\ }\href@noop {} {\bibfield  {journal} {\bibinfo  {journal} {Mem.
  Am. Math. Soc.}\ }\textbf {\bibinfo {volume} {123}},\ \bibinfo {pages}
  {x+110} (\bibinfo {year} {1996})}\BibitemShut {NoStop}%
\bibitem [{\citenamefont {{Arn\'{e}odo}}\ \emph
  {et~al.}(1998{\natexlab{a}})\citenamefont {{Arn\'{e}odo}}, \citenamefont
  {{Bacry}}, \citenamefont {{Jaffard}},\ and\ \citenamefont
  {{Muzy}}}]{arneodo98}%
  \BibitemOpen
  \bibfield  {author} {\bibinfo {author} {\bibfnamefont {A.}~\bibnamefont
  {{Arn\'{e}odo}}}, \bibinfo {author} {\bibfnamefont {E.}~\bibnamefont
  {{Bacry}}}, \bibinfo {author} {\bibfnamefont {S.}~\bibnamefont {{Jaffard}}},
  \ and\ \bibinfo {author} {\bibfnamefont {J.~F.}\ \bibnamefont {{Muzy}}},\
  }\href@noop {} {\bibfield  {journal} {\bibinfo  {journal} {J. Fourier Analys.
  Appl.}\ }\textbf {\bibinfo {volume} {4}},\ \bibinfo {pages} {159} (\bibinfo
  {year} {1998}{\natexlab{a}})}\BibitemShut {NoStop}%
\bibitem [{\citenamefont {{Hunt}}\ and\ \citenamefont
  {{Vassilicos}}(1991)}]{hunt91}%
  \BibitemOpen
  \bibfield  {author} {\bibinfo {author} {\bibfnamefont {J.~C.~R.}\
  \bibnamefont {{Hunt}}}\ and\ \bibinfo {author} {\bibfnamefont {J.~C.}\
  \bibnamefont {{Vassilicos}}},\ }\href@noop {} {\bibfield  {journal} {\bibinfo
   {journal} {Proc. Roy. Soc. Lond. Ser. A.}\ }\textbf {\bibinfo {volume}
  {435}},\ \bibinfo {pages} {505} (\bibinfo {year} {1991})}\BibitemShut
  {NoStop}%
\bibitem [{\citenamefont {{Nicolleau}}\ and\ \citenamefont
  {{Vassilicos}}(1999)}]{nicolleau99}%
  \BibitemOpen
  \bibfield  {author} {\bibinfo {author} {\bibfnamefont {F.}~\bibnamefont
  {{Nicolleau}}}\ and\ \bibinfo {author} {\bibfnamefont {J.~C.}\ \bibnamefont
  {{Vassilicos}}},\ }\href@noop {} {\bibfield  {journal} {\bibinfo  {journal}
  {Phil. Trans. R. Soc. Lond., Ser. A}\ }\textbf {\bibinfo {volume} {357}},\
  \bibinfo {pages} {2439} (\bibinfo {year} {1999})}\BibitemShut {NoStop}%
\bibitem [{\citenamefont {{Kolmogorov}}(1962)}]{K62}%
  \BibitemOpen
  \bibfield  {author} {\bibinfo {author} {\bibfnamefont {A.~N.}\ \bibnamefont
  {{Kolmogorov}}},\ }\href@noop {} {\bibfield  {journal} {\bibinfo  {journal}
  {J. Fluid Mech.}\ }\textbf {\bibinfo {volume} {13}},\ \bibinfo {pages} {82}
  (\bibinfo {year} {1962})}\BibitemShut {NoStop}%
\bibitem [{\citenamefont {{Praskovsky}}\ \emph {et~al.}(1993)\citenamefont
  {{Praskovsky}}, \citenamefont {{Gledzer}}, \citenamefont {{Karyakin}},\ and\
  \citenamefont {{Zhou}}}]{prask93}%
  \BibitemOpen
  \bibfield  {author} {\bibinfo {author} {\bibfnamefont {A.~A.}\ \bibnamefont
  {{Praskovsky}}}, \bibinfo {author} {\bibfnamefont {E.~B.}\ \bibnamefont
  {{Gledzer}}}, \bibinfo {author} {\bibfnamefont {M.~Y.}\ \bibnamefont
  {{Karyakin}}}, \ and\ \bibinfo {author} {\bibfnamefont {Y.}~\bibnamefont
  {{Zhou}}},\ }\href@noop {} {\bibfield  {journal} {\bibinfo  {journal} {J.
  Fluid Mech.}\ }\textbf {\bibinfo {volume} {248}},\ \bibinfo {pages} {493}
  (\bibinfo {year} {1993})}\BibitemShut {NoStop}%
\bibitem [{\citenamefont {{Stresing}}\ and\ \citenamefont
  {{Peinke}}(2010)}]{stresing10}%
  \BibitemOpen
  \bibfield  {author} {\bibinfo {author} {\bibfnamefont {R.}~\bibnamefont
  {{Stresing}}}\ and\ \bibinfo {author} {\bibfnamefont {J.}~\bibnamefont
  {{Peinke}}},\ }\href {\doibase 10.1088/1367-2630/12/10/103046} {\bibfield
  {journal} {\bibinfo  {journal} {New J. Phys.}\ }\textbf {\bibinfo {volume}
  {12}},\ \bibinfo {pages} {103046} (\bibinfo {year} {2010})}\BibitemShut
  {NoStop}%
\bibitem [{\citenamefont {{Keylock}}\ \emph
  {et~al.}(2012{\natexlab{a}})\citenamefont {{Keylock}}, \citenamefont
  {{Nishimura}},\ and\ \citenamefont {{Peinke}}}]{k12}%
  \BibitemOpen
  \bibfield  {author} {\bibinfo {author} {\bibfnamefont {C.~J.}\ \bibnamefont
  {{Keylock}}}, \bibinfo {author} {\bibfnamefont {K.}~\bibnamefont
  {{Nishimura}}}, \ and\ \bibinfo {author} {\bibfnamefont {J.}~\bibnamefont
  {{Peinke}}},\ }\href {\doibase 10.1029/2011JF002127} {\bibfield  {journal}
  {\bibinfo  {journal} {J. Geophys. Res.}\ }\textbf {\bibinfo {volume} {117}}
  (\bibinfo {year} {2012}{\natexlab{a}}),\ 10.1029/2011JF002127}\BibitemShut
  {NoStop}%
\bibitem [{\citenamefont {{Frisch}}\ and\ \citenamefont
  {{Parisi}}(1985)}]{fp85}%
  \BibitemOpen
  \bibfield  {author} {\bibinfo {author} {\bibfnamefont {U.}~\bibnamefont
  {{Frisch}}}\ and\ \bibinfo {author} {\bibfnamefont {G.}~\bibnamefont
  {{Parisi}}},\ }in\ \href@noop {} {\emph {\bibinfo {booktitle} {Turbulence and
  Predictability in Geophysical Fluid Dynamics and Climate Dynamics}}},\
  \bibinfo {editor} {edited by\ \bibinfo {editor} {\bibfnamefont
  {M.}~\bibnamefont {{Ghil}}}, \bibinfo {editor} {\bibfnamefont
  {R.}~\bibnamefont {{Benzi}}}, \ and\ \bibinfo {editor} {\bibfnamefont
  {G.}~\bibnamefont {{Parisi}}}}\ (\bibinfo  {publisher} {North-Holland,
  Amsterdam},\ \bibinfo {year} {1985})\ pp.\ \bibinfo {pages}
  {84--88}\BibitemShut {NoStop}%
\bibitem [{\citenamefont {{Benzi}}\ \emph
  {et~al.}(1993{\natexlab{a}})\citenamefont {{Benzi}}, \citenamefont
  {{Ciliberto}}, \citenamefont {{Tripiccione}}, \citenamefont {{Baudet}},\ and\
  \citenamefont {{Succi}}}]{ess}%
  \BibitemOpen
  \bibfield  {author} {\bibinfo {author} {\bibfnamefont {R.}~\bibnamefont
  {{Benzi}}}, \bibinfo {author} {\bibfnamefont {S.}~\bibnamefont
  {{Ciliberto}}}, \bibinfo {author} {\bibfnamefont {R.}~\bibnamefont
  {{Tripiccione}}}, \bibinfo {author} {\bibfnamefont {C.}~\bibnamefont
  {{Baudet}}}, \ and\ \bibinfo {author} {\bibfnamefont {S.}~\bibnamefont
  {{Succi}}},\ }\href@noop {} {\bibfield  {journal} {\bibinfo  {journal} {Phys.
  Rev E}\ }\textbf {\bibinfo {volume} {48}},\ \bibinfo {pages} {29} (\bibinfo
  {year} {1993}{\natexlab{a}})}\BibitemShut {NoStop}%
\bibitem [{\citenamefont {{Schertzer}}\ and\ \citenamefont
  {{Lovejoy}}(1987)}]{schertzer87}%
  \BibitemOpen
  \bibfield  {author} {\bibinfo {author} {\bibfnamefont {D.}~\bibnamefont
  {{Schertzer}}}\ and\ \bibinfo {author} {\bibfnamefont {S.}~\bibnamefont
  {{Lovejoy}}},\ }\href@noop {} {\bibfield  {journal} {\bibinfo  {journal} {J.
  Geophys. Res.}\ }\textbf {\bibinfo {volume} {92}},\ \bibinfo {pages} {9693}
  (\bibinfo {year} {1987})}\BibitemShut {NoStop}%
\bibitem [{\citenamefont {{Muzy}}\ \emph {et~al.}(1991)\citenamefont {{Muzy}},
  \citenamefont {{Bacry}},\ and\ \citenamefont {Arn\'{e}odo}}]{MUZY91}%
  \BibitemOpen
  \bibfield  {author} {\bibinfo {author} {\bibfnamefont {J.~F.}\ \bibnamefont
  {{Muzy}}}, \bibinfo {author} {\bibfnamefont {E.}~\bibnamefont {{Bacry}}}, \
  and\ \bibinfo {author} {\bibfnamefont {A.}~\bibnamefont {Arn\'{e}odo}},\
  }\href@noop {} {\bibfield  {journal} {\bibinfo  {journal} {Phys. Rev. Lett.}\
  }\textbf {\bibinfo {volume} {67}},\ \bibinfo {pages} {3515} (\bibinfo {year}
  {1991})}\BibitemShut {NoStop}%
\bibitem [{\citenamefont {{Bacry}}\ \emph {et~al.}(1993)\citenamefont
  {{Bacry}}, \citenamefont {{Muzy}},\ and\ \citenamefont
  {{Arn\'{e}odo}}}]{BMA93}%
  \BibitemOpen
  \bibfield  {author} {\bibinfo {author} {\bibfnamefont {E.}~\bibnamefont
  {{Bacry}}}, \bibinfo {author} {\bibfnamefont {J.}~\bibnamefont {{Muzy}}}, \
  and\ \bibinfo {author} {\bibfnamefont {A.}~\bibnamefont {{Arn\'{e}odo}}},\
  }\href@noop {} {\bibfield  {journal} {\bibinfo  {journal} {J. Stat. Phys.}\
  }\textbf {\bibinfo {volume} {70}},\ \bibinfo {pages} {635} (\bibinfo {year}
  {1993})}\BibitemShut {NoStop}%
\bibitem [{\citenamefont {{Seuret}}\ and\ \citenamefont {{L\'{e}vy
  V\'{e}hel}}(2002)}]{SL02}%
  \BibitemOpen
  \bibfield  {author} {\bibinfo {author} {\bibfnamefont {S.}~\bibnamefont
  {{Seuret}}}\ and\ \bibinfo {author} {\bibfnamefont {J.}~\bibnamefont
  {{L\'{e}vy V\'{e}hel}}},\ }\href@noop {} {\bibfield  {journal} {\bibinfo
  {journal} {Appl. Comput. Harmonic Anal.}\ }\textbf {\bibinfo {volume} {13}},\
  \bibinfo {pages} {263} (\bibinfo {year} {2002})}\BibitemShut {NoStop}%
\bibitem [{\citenamefont {{Kolwankar}}\ and\ \citenamefont {{L\'{e}vy
  V\'{e}hel}}(2002)}]{KL02}%
  \BibitemOpen
  \bibfield  {author} {\bibinfo {author} {\bibfnamefont {K.}~\bibnamefont
  {{Kolwankar}}}\ and\ \bibinfo {author} {\bibfnamefont {J.}~\bibnamefont
  {{L\'{e}vy V\'{e}hel}}},\ }\href@noop {} {\bibfield  {journal} {\bibinfo
  {journal} {J. Fourier Anal. Appl.}\ }\textbf {\bibinfo {volume} {8}},\
  \bibinfo {pages} {319} (\bibinfo {year} {2002})}\BibitemShut {NoStop}%
\bibitem [{\citenamefont {{Seuret}}\ and\ \citenamefont {{L\'{e}vy
  V\'{e}hel}}(2003)}]{SLV03}%
  \BibitemOpen
  \bibfield  {author} {\bibinfo {author} {\bibfnamefont {S.}~\bibnamefont
  {{Seuret}}}\ and\ \bibinfo {author} {\bibfnamefont {J.}~\bibnamefont
  {{L\'{e}vy V\'{e}hel}}},\ }\href@noop {} {\bibfield  {journal} {\bibinfo
  {journal} {J. Fourier Anal. Appl.}\ }\textbf {\bibinfo {volume} {9}},\
  \bibinfo {pages} {473} (\bibinfo {year} {2003})}\BibitemShut {NoStop}%
\bibitem [{\citenamefont {INRIA}(2014)}]{fraclab}%
  \BibitemOpen
  \bibfield  {author} {\bibinfo {author} {\bibnamefont {INRIA}},\ }\href
  {http://fraclab.saclay.inria.fr/} {\enquote {\bibinfo {title} {The fraclab
  toolbox (http://fraclab.saclay.inria.fr/)},}\ } (\bibinfo {year}
  {2014})\BibitemShut {NoStop}%
\bibitem [{\citenamefont {{Benzi}}\ \emph
  {et~al.}(1993{\natexlab{b}})\citenamefont {{Benzi}}, \citenamefont
  {{Biferale}}, \citenamefont {{Crisanti}}, \citenamefont {{Paladin}},
  \citenamefont {{Vergassola}},\ and\ \citenamefont {{Vulpiani}}}]{B93}%
  \BibitemOpen
  \bibfield  {author} {\bibinfo {author} {\bibfnamefont {R.}~\bibnamefont
  {{Benzi}}}, \bibinfo {author} {\bibfnamefont {L.}~\bibnamefont {{Biferale}}},
  \bibinfo {author} {\bibfnamefont {A.}~\bibnamefont {{Crisanti}}}, \bibinfo
  {author} {\bibfnamefont {G.}~\bibnamefont {{Paladin}}}, \bibinfo {author}
  {\bibfnamefont {M.}~\bibnamefont {{Vergassola}}}, \ and\ \bibinfo {author}
  {\bibfnamefont {A.}~\bibnamefont {{Vulpiani}}},\ }\href@noop {} {\bibfield
  {journal} {\bibinfo  {journal} {Physica D}\ }\textbf {\bibinfo {volume}
  {65}},\ \bibinfo {pages} {352} (\bibinfo {year}
  {1993}{\natexlab{b}})}\BibitemShut {NoStop}%
\bibitem [{\citenamefont {{Arn\'{e}odo}}\ \emph
  {et~al.}(1998{\natexlab{b}})\citenamefont {{Arn\'{e}odo}}, \citenamefont
  {{Bacry}},\ and\ \citenamefont {{Muzy}}}]{ABM98}%
  \BibitemOpen
  \bibfield  {author} {\bibinfo {author} {\bibfnamefont {A.}~\bibnamefont
  {{Arn\'{e}odo}}}, \bibinfo {author} {\bibfnamefont {E.}~\bibnamefont
  {{Bacry}}}, \ and\ \bibinfo {author} {\bibfnamefont {J.~F.}\ \bibnamefont
  {{Muzy}}},\ }\href@noop {} {\bibfield  {journal} {\bibinfo  {journal} {J.
  Math. Phys.}\ }\textbf {\bibinfo {volume} {39}},\ \bibinfo {pages} {4142}
  (\bibinfo {year} {1998}{\natexlab{b}})}\BibitemShut {NoStop}%
\bibitem [{\citenamefont {{Benassi}}\ \emph {et~al.}(1997)\citenamefont
  {{Benassi}}, \citenamefont {{Jaffard}},\ and\ \citenamefont
  {{Roux}}}]{benassi97}%
  \BibitemOpen
  \bibfield  {author} {\bibinfo {author} {\bibfnamefont {A.}~\bibnamefont
  {{Benassi}}}, \bibinfo {author} {\bibfnamefont {S.}~\bibnamefont
  {{Jaffard}}}, \ and\ \bibinfo {author} {\bibfnamefont {D.}~\bibnamefont
  {{Roux}}},\ }\href@noop {} {\bibfield  {journal} {\bibinfo  {journal} {Rev.
  Mat. Iberoamericana}\ }\textbf {\bibinfo {volume} {13}},\ \bibinfo {pages}
  {19–89} (\bibinfo {year} {1997})}\BibitemShut {NoStop}%
\bibitem [{\citenamefont {{Peltier}}\ and\ \citenamefont {{L\'{e}vy
  V\'{e}hel}}(1995)}]{peltier95}%
  \BibitemOpen
  \bibfield  {author} {\bibinfo {author} {\bibfnamefont {R.}~\bibnamefont
  {{Peltier}}}\ and\ \bibinfo {author} {\bibfnamefont {J.}~\bibnamefont
  {{L\'{e}vy V\'{e}hel}}},\ }\href@noop {} {\emph {\bibinfo {title}
  {Multifractional Brownian Motion: Definition and Preliminary Results}}},\
  \bibinfo {type} {Research report}\ \bibinfo {number} {2645}\ (\bibinfo
  {institution} {INRIA},\ \bibinfo {year} {1995})\BibitemShut {NoStop}%
\bibitem [{\citenamefont {{Wood}}\ and\ \citenamefont
  {{Chan}}(1994)}]{woodchan}%
  \BibitemOpen
  \bibfield  {author} {\bibinfo {author} {\bibfnamefont {A.}~\bibnamefont
  {{Wood}}}\ and\ \bibinfo {author} {\bibfnamefont {G.}~\bibnamefont
  {{Chan}}},\ }\href@noop {} {\bibfield  {journal} {\bibinfo  {journal} {J.
  Comput. Graph. Statist.}\ }\textbf {\bibinfo {volume} {3}},\ \bibinfo {pages}
  {409–432} (\bibinfo {year} {1994})}\BibitemShut {NoStop}%
\bibitem [{\citenamefont {{Selesnick}}(2002)}]{seles02}%
  \BibitemOpen
  \bibfield  {author} {\bibinfo {author} {\bibfnamefont {I.}~\bibnamefont
  {{Selesnick}}},\ }\href@noop {} {\bibfield  {journal} {\bibinfo  {journal}
  {IEEE Trans. Sig. Proc.}\ }\textbf {\bibinfo {volume} {50}},\ \bibinfo
  {pages} {1144} (\bibinfo {year} {2002})}\BibitemShut {NoStop}%
\bibitem [{\citenamefont {{Abry}}\ and\ \citenamefont
  {{Flandrin}}(1994)}]{af94}%
  \BibitemOpen
  \bibfield  {author} {\bibinfo {author} {\bibfnamefont {P.}~\bibnamefont
  {{Abry}}}\ and\ \bibinfo {author} {\bibfnamefont {P.}~\bibnamefont
  {{Flandrin}}},\ }in\ \href@noop {} {\emph {\bibinfo {booktitle} {Proc.
  IEEE-SP Int. Symp. Time-Freq. Time-Scale Anal.}}}\ (\bibinfo {year} {1994})\
  pp.\ \bibinfo {pages} {225--228}\BibitemShut {NoStop}%
\bibitem [{\citenamefont {{Ozturk}}\ \emph {et~al.}(2000)\citenamefont
  {{Ozturk}}, \citenamefont {{Kucur}},\ and\ \citenamefont {{Atkin}}}]{ozturk}%
  \BibitemOpen
  \bibfield  {author} {\bibinfo {author} {\bibfnamefont {E.}~\bibnamefont
  {{Ozturk}}}, \bibinfo {author} {\bibfnamefont {O.}~\bibnamefont {{Kucur}}}, \
  and\ \bibinfo {author} {\bibfnamefont {G.}~\bibnamefont {{Atkin}}},\ }in\
  \href@noop {} {\emph {\bibinfo {booktitle} {Proc. IEEE Int. Conf. Acoust.,
  Speech, Signal Process.}}}\ (\bibinfo {year} {2000})\ p.\ \bibinfo {pages}
  {2641}\BibitemShut {NoStop}%
\bibitem [{\citenamefont {{Kingsbury}}(2001)}]{kings01}%
  \BibitemOpen
  \bibfield  {author} {\bibinfo {author} {\bibfnamefont {N.}~\bibnamefont
  {{Kingsbury}}},\ }\href@noop {} {\bibfield  {journal} {\bibinfo  {journal}
  {Appl. Comput. Harmon. Anal.}\ }\textbf {\bibinfo {volume} {10}},\ \bibinfo
  {pages} {234} (\bibinfo {year} {2001})}\BibitemShut {NoStop}%
\bibitem [{\citenamefont {{Selesnick}}\ \emph {et~al.}(2005)\citenamefont
  {{Selesnick}}, \citenamefont {{Baraniuk}},\ and\ \citenamefont
  {{Kingsbury}}}]{seles05}%
  \BibitemOpen
  \bibfield  {author} {\bibinfo {author} {\bibfnamefont {I.}~\bibnamefont
  {{Selesnick}}}, \bibinfo {author} {\bibfnamefont {R.}~\bibnamefont
  {{Baraniuk}}}, \ and\ \bibinfo {author} {\bibfnamefont {N.}~\bibnamefont
  {{Kingsbury}}},\ }\href@noop {} {\bibfield  {journal} {\bibinfo  {journal}
  {IEEE Signal Proc. Mag.}\ }\textbf {\bibinfo {volume} {22}},\ \bibinfo
  {pages} {123} (\bibinfo {year} {2005})}\BibitemShut {NoStop}%
\bibitem [{\citenamefont {{Selesnick}}(2001)}]{seles01}%
  \BibitemOpen
  \bibfield  {author} {\bibinfo {author} {\bibfnamefont {I.}~\bibnamefont
  {{Selesnick}}},\ }\href@noop {} {\bibfield  {journal} {\bibinfo  {journal}
  {IEEE Sig. Proc. Lett.}\ }\textbf {\bibinfo {volume} {8}},\ \bibinfo {pages}
  {170} (\bibinfo {year} {2001})}\BibitemShut {NoStop}%
\bibitem [{\citenamefont {{Liang}}\ and\ \citenamefont {{Parks}}(1996)}]{lp96}%
  \BibitemOpen
  \bibfield  {author} {\bibinfo {author} {\bibfnamefont {J.}~\bibnamefont
  {{Liang}}}\ and\ \bibinfo {author} {\bibfnamefont {T.~W.}\ \bibnamefont
  {{Parks}}},\ }\href@noop {} {\bibfield  {journal} {\bibinfo  {journal} {IEEE
  T. Signal Proces.}\ }\textbf {\bibinfo {volume} {44}},\ \bibinfo {pages}
  {225} (\bibinfo {year} {1996})}\BibitemShut {NoStop}%
\bibitem [{\citenamefont {{Falconer}}(1994)}]{falconer}%
  \BibitemOpen
  \bibfield  {author} {\bibinfo {author} {\bibfnamefont {K.}~\bibnamefont
  {{Falconer}}},\ }\href@noop {} {\emph {\bibinfo {title} {The Geometry of
  Fractal Sets}}}\ (\bibinfo  {publisher} {OUP},\ \bibinfo {address} {Oxford,
  U.K.},\ \bibinfo {year} {1994})\BibitemShut {NoStop}%
\bibitem [{\citenamefont {{Keylock}}\ \emph
  {et~al.}(2012{\natexlab{b}})\citenamefont {{Keylock}}, \citenamefont
  {{Nishimura}}, \citenamefont {{Nemoto}},\ and\ \citenamefont {{Ito}}}]{k12b}%
  \BibitemOpen
  \bibfield  {author} {\bibinfo {author} {\bibfnamefont {C.~J.}\ \bibnamefont
  {{Keylock}}}, \bibinfo {author} {\bibfnamefont {K.}~\bibnamefont
  {{Nishimura}}}, \bibinfo {author} {\bibfnamefont {M.}~\bibnamefont
  {{Nemoto}}}, \ and\ \bibinfo {author} {\bibfnamefont {Y.}~\bibnamefont
  {{Ito}}},\ }\href {\doibase 10.1007/s10652-011-9233-0} {\bibfield  {journal}
  {\bibinfo  {journal} {Environ. Fluid Mech.}\ } (\bibinfo {year}
  {2012}{\natexlab{b}}),\ 10.1007/s10652-011-9233-0}\BibitemShut {NoStop}%
\bibitem [{\citenamefont {{Moffat}}(1984)}]{moffat84}%
  \BibitemOpen
  \bibfield  {author} {\bibinfo {author} {\bibfnamefont {K.}~\bibnamefont
  {{Moffat}}},\ }in\ \href@noop {} {\emph {\bibinfo {booktitle} {Proc. IUTAM
  Symp. on Turbulence and Chaotic Phenomena in Fluids}}},\ \bibinfo {editor}
  {edited by\ \bibinfo {editor} {\bibfnamefont {T.}~\bibnamefont {{Tatsumi}}}}\
  (\bibinfo  {publisher} {Elsevier},\ \bibinfo {year} {1984})\BibitemShut
  {NoStop}%
\bibitem [{\citenamefont {{Ide}}\ and\ \citenamefont
  {{Sornette}}(2002)}]{sornette02}%
  \BibitemOpen
  \bibfield  {author} {\bibinfo {author} {\bibfnamefont {K.}~\bibnamefont
  {{Ide}}}\ and\ \bibinfo {author} {\bibfnamefont {D.}~\bibnamefont
  {{Sornette}}},\ }\href@noop {} {\bibfield  {journal} {\bibinfo  {journal}
  {Physica A}\ }\textbf {\bibinfo {volume} {307}},\ \bibinfo {pages} {63}
  (\bibinfo {year} {2002})}\BibitemShut {NoStop}%
\bibitem [{\citenamefont {{Zhou}}\ and\ \citenamefont
  {{Sornette}}(2003)}]{sornette03}%
  \BibitemOpen
  \bibfield  {author} {\bibinfo {author} {\bibfnamefont {W.~X.}\ \bibnamefont
  {{Zhou}}}\ and\ \bibinfo {author} {\bibfnamefont {D.}~\bibnamefont
  {{Sornette}}},\ }\href {\doibase 10.1016/j.physa.2002.12.001} {\bibfield
  {journal} {\bibinfo  {journal} {Physica A}\ }\textbf {\bibinfo {volume}
  {330}},\ \bibinfo {pages} {543} (\bibinfo {year} {2003})}\BibitemShut
  {NoStop}%
\bibitem [{\citenamefont {{Seuret}}(2006)}]{seuret06}%
  \BibitemOpen
  \bibfield  {author} {\bibinfo {author} {\bibfnamefont {S.}~\bibnamefont
  {{Seuret}}},\ }\href {\doibase 10.1002/mana.200510417} {\bibfield  {journal}
  {\bibinfo  {journal} {Math. Nachr.}\ }\textbf {\bibinfo {volume} {279}},\
  \bibinfo {pages} {1195} (\bibinfo {year} {2006})}\BibitemShut {NoStop}%
\bibitem [{\citenamefont {{Aubry}}\ and\ \citenamefont
  {{Jaffard}}(2002)}]{aubry02}%
  \BibitemOpen
  \bibfield  {author} {\bibinfo {author} {\bibfnamefont {J.~M.}\ \bibnamefont
  {{Aubry}}}\ and\ \bibinfo {author} {\bibfnamefont {S.}~\bibnamefont
  {{Jaffard}}},\ }\href@noop {} {\bibfield  {journal} {\bibinfo  {journal}
  {Comm. Math. Phys.}\ }\textbf {\bibinfo {volume} {227}},\ \bibinfo {pages}
  {483} (\bibinfo {year} {2002})}\BibitemShut {NoStop}%
\bibitem [{\citenamefont {{Donoho}}\ and\ \citenamefont
  {{Johnstone}}(1994)}]{DJ94}%
  \BibitemOpen
  \bibfield  {author} {\bibinfo {author} {\bibfnamefont {D.~L.}\ \bibnamefont
  {{Donoho}}}\ and\ \bibinfo {author} {\bibfnamefont {I.~M.}\ \bibnamefont
  {{Johnstone}}},\ }\href@noop {} {\bibfield  {journal} {\bibinfo  {journal}
  {Biometrika}\ }\textbf {\bibinfo {volume} {81}},\ \bibinfo {pages} {425}
  (\bibinfo {year} {1994})}\BibitemShut {NoStop}%
\bibitem [{\citenamefont {{Oboukhov}}(1962)}]{Oboukhov62}%
  \BibitemOpen
  \bibfield  {author} {\bibinfo {author} {\bibfnamefont {A.~M.}\ \bibnamefont
  {{Oboukhov}}},\ }\href@noop {} {\bibfield  {journal} {\bibinfo  {journal} {J.
  Fluid Mech.}\ }\textbf {\bibinfo {volume} {13}},\ \bibinfo {pages} {77}
  (\bibinfo {year} {1962})}\BibitemShut {NoStop}%
\bibitem [{\citenamefont {{Kolmogorov}}(1941)}]{K41}%
  \BibitemOpen
  \bibfield  {author} {\bibinfo {author} {\bibfnamefont {A.~N.}\ \bibnamefont
  {{Kolmogorov}}},\ }\href@noop {} {\bibfield  {journal} {\bibinfo  {journal}
  {Dokl. Akad. Nauk. SSSR.}\ }\textbf {\bibinfo {volume} {30}},\ \bibinfo
  {pages} {299} (\bibinfo {year} {1941})}\BibitemShut {NoStop}%
\bibitem [{\citenamefont {{Frisch}}\ \emph {et~al.}(1978)\citenamefont
  {{Frisch}}, \citenamefont {{Sulem}},\ and\ \citenamefont
  {{Nelkin}}}]{frisch78}%
  \BibitemOpen
  \bibfield  {author} {\bibinfo {author} {\bibfnamefont {U.}~\bibnamefont
  {{Frisch}}}, \bibinfo {author} {\bibfnamefont {P.~L.}\ \bibnamefont
  {{Sulem}}}, \ and\ \bibinfo {author} {\bibfnamefont {M.}~\bibnamefont
  {{Nelkin}}},\ }\href@noop {} {\bibfield  {journal} {\bibinfo  {journal} {J.
  Fluid Mech.}\ }\textbf {\bibinfo {volume} {87}},\ \bibinfo {pages} {719}
  (\bibinfo {year} {1978})}\BibitemShut {NoStop}%
\bibitem [{\citenamefont {{Meneveau}}\ and\ \citenamefont
  {{Sreenivasan}}(1987)}]{MS87}%
  \BibitemOpen
  \bibfield  {author} {\bibinfo {author} {\bibfnamefont {C.}~\bibnamefont
  {{Meneveau}}}\ and\ \bibinfo {author} {\bibfnamefont {K.}~\bibnamefont
  {{Sreenivasan}}},\ }\href@noop {} {\bibfield  {journal} {\bibinfo  {journal}
  {Phys. Rev. Lett.}\ }\textbf {\bibinfo {volume} {59}},\ \bibinfo {pages}
  {1424} (\bibinfo {year} {1987})}\BibitemShut {NoStop}%
\bibitem [{\citenamefont {{Frisch}}\ and\ \citenamefont
  {{Vergassola}}(1991)}]{FV91}%
  \BibitemOpen
  \bibfield  {author} {\bibinfo {author} {\bibfnamefont {U.}~\bibnamefont
  {{Frisch}}}\ and\ \bibinfo {author} {\bibfnamefont {M.}~\bibnamefont
  {{Vergassola}}},\ }\href@noop {} {\bibfield  {journal} {\bibinfo  {journal}
  {Europhys. Lett.}\ }\textbf {\bibinfo {volume} {14}},\ \bibinfo {pages} {439}
  (\bibinfo {year} {1991})}\BibitemShut {NoStop}%
\bibitem [{\citenamefont {{Frisch}}\ \emph {et~al.}(2005)\citenamefont
  {{Frisch}}, \citenamefont {{Bec}},\ and\ \citenamefont
  {{Aurell}}}]{frisch05}%
  \BibitemOpen
  \bibfield  {author} {\bibinfo {author} {\bibfnamefont {U.}~\bibnamefont
  {{Frisch}}}, \bibinfo {author} {\bibfnamefont {J.}~\bibnamefont {{Bec}}}, \
  and\ \bibinfo {author} {\bibfnamefont {E.}~\bibnamefont {{Aurell}}},\ }\href
  {\doibase 10.1063/1.2008994} {\bibfield  {journal} {\bibinfo  {journal}
  {Phys. Fluids}\ }\textbf {\bibinfo {volume} {17}} (\bibinfo {year} {2005}),\
  10.1063/1.2008994}\BibitemShut {NoStop}%
\bibitem [{\citenamefont {{Nakagawa}}\ and\ \citenamefont
  {{Nezu}}(1977)}]{nezu77}%
  \BibitemOpen
  \bibfield  {author} {\bibinfo {author} {\bibfnamefont {H.}~\bibnamefont
  {{Nakagawa}}}\ and\ \bibinfo {author} {\bibfnamefont {I.}~\bibnamefont
  {{Nezu}}},\ }\href@noop {} {\bibfield  {journal} {\bibinfo  {journal} {J.
  Fluid Mech.}\ }\textbf {\bibinfo {volume} {80}},\ \bibinfo {pages} {99}
  (\bibinfo {year} {1977})}\BibitemShut {NoStop}%
\bibitem [{\citenamefont {{Bogard}}\ and\ \citenamefont
  {{Tiederman}}(1986)}]{bt86}%
  \BibitemOpen
  \bibfield  {author} {\bibinfo {author} {\bibfnamefont {D.~G.}\ \bibnamefont
  {{Bogard}}}\ and\ \bibinfo {author} {\bibfnamefont {W.~G.}\ \bibnamefont
  {{Tiederman}}},\ }\href@noop {} {\bibfield  {journal} {\bibinfo  {journal}
  {J. Fluid Mech.}\ }\textbf {\bibinfo {volume} {162}},\ \bibinfo {pages} {389}
  (\bibinfo {year} {1986})}\BibitemShut {NoStop}%
\bibitem [{\citenamefont {{Keylock}}\ \emph {et~al.}(2013)\citenamefont
  {{Keylock}}, \citenamefont {{Singh}},\ and\ \citenamefont
  {{Foufoula-Georgiou}}}]{k13}%
  \BibitemOpen
  \bibfield  {author} {\bibinfo {author} {\bibfnamefont {C.~J.}\ \bibnamefont
  {{Keylock}}}, \bibinfo {author} {\bibfnamefont {A.}~\bibnamefont {{Singh}}},
  \ and\ \bibinfo {author} {\bibfnamefont {E.}~\bibnamefont
  {{Foufoula-Georgiou}}},\ }\href@noop {} {\bibfield  {journal} {\bibinfo
  {journal} {Geophys. Res. Lett.}\ }\textbf {\bibinfo {volume} {40}},\ \bibinfo
  {pages} {doi:10.1002/grl.50337} (\bibinfo {year} {2013})}\BibitemShut
  {NoStop}%
\bibitem [{\citenamefont {{Keylock}}\ \emph {et~al.}(2016)\citenamefont
  {{Keylock}}, \citenamefont {{Chang}},\ and\ \citenamefont
  {{Constantinescu}}}]{k16}%
  \BibitemOpen
  \bibfield  {author} {\bibinfo {author} {\bibfnamefont {C.~J.}\ \bibnamefont
  {{Keylock}}}, \bibinfo {author} {\bibfnamefont {K.~S.}\ \bibnamefont
  {{Chang}}}, \ and\ \bibinfo {author} {\bibfnamefont {G.~S.}\ \bibnamefont
  {{Constantinescu}}},\ }\href@noop {} {\bibfield  {journal} {\bibinfo
  {journal} {J. Fluid Mech.}\ }\textbf {\bibinfo {volume} {805}},\ \bibinfo
  {pages} {656} (\bibinfo {year} {2016})}\BibitemShut {NoStop}%
\bibitem [{\citenamefont {{Renner}}\ \emph {et~al.}(2001)\citenamefont
  {{Renner}}, \citenamefont {{Peinke}},\ and\ \citenamefont
  {{Friedrich}}}]{renner01}%
  \BibitemOpen
  \bibfield  {author} {\bibinfo {author} {\bibfnamefont {C.}~\bibnamefont
  {{Renner}}}, \bibinfo {author} {\bibfnamefont {J.}~\bibnamefont {{Peinke}}},
  \ and\ \bibinfo {author} {\bibfnamefont {R.}~\bibnamefont {{Friedrich}}},\
  }\href@noop {} {\bibfield  {journal} {\bibinfo  {journal} {J. Fluid Mech.}\
  }\textbf {\bibinfo {volume} {433}},\ \bibinfo {pages} {383} (\bibinfo {year}
  {2001})}\BibitemShut {NoStop}%
\bibitem [{\citenamefont {{L\'{e}vy V\'{e}hel}}(2013)}]{levyvehel13}%
  \BibitemOpen
  \bibfield  {author} {\bibinfo {author} {\bibfnamefont {J.}~\bibnamefont
  {{L\'{e}vy V\'{e}hel}}},\ }\href@noop {} {\bibfield  {journal} {\bibinfo
  {journal} {Nonlin. Proc. Geophys.}\ }\textbf {\bibinfo {volume} {20}},\
  \bibinfo {pages} {643} (\bibinfo {year} {2013})}\BibitemShut {NoStop}%
\bibitem [{\citenamefont {{Echelard}}\ \emph {et~al.}(2015)\citenamefont
  {{Echelard}}, \citenamefont {{L\'{e}vy V\'{e}hel}},\ and\ \citenamefont
  {{Philippe}}}]{echelard15}%
  \BibitemOpen
  \bibfield  {author} {\bibinfo {author} {\bibfnamefont {A.}~\bibnamefont
  {{Echelard}}}, \bibinfo {author} {\bibfnamefont {J.}~\bibnamefont {{L\'{e}vy
  V\'{e}hel}}}, \ and\ \bibinfo {author} {\bibfnamefont {A.}~\bibnamefont
  {{Philippe}}},\ }\href@noop {} {\bibfield  {journal} {\bibinfo  {journal}
  {Scand. J. Stat.}\ }\textbf {\bibinfo {volume} {42}},\ \bibinfo {pages} {485}
  (\bibinfo {year} {2015})}\BibitemShut {NoStop}%
\end{thebibliography}%

\end{document}